\documentstyle[aps]{revtex}

\begin{document}
\author{Kaige Wang}
\address{CCAST (World Laboratory), P. O. Box 8730, Beijing 100080, and \\
Department of Physics, Applied Optics Beijing Area Major Laboratory, Beijing%
\\
Normal University, Beijing 100875, China}
\title{Quantum Theory of Two-Photon Wavepacket Interference in a Beam Splitter }
\maketitle

\begin{abstract}
We study a general theory on the interference of a two-photon wavepacket in
a beam splitter. The theory is carried out in the Schr\"{o}dinger picture so
that the quantum nature of the two-photon interference is explicitly
presented. We find that the topological symmetry of the
probability-amplitude spectrum of the two-photon wavepacket dominates the
manners of two-photon interference which are distinguished according to
increasing and decreasing the coincidence probability for the absence of
interference. However, two-photon entanglement can be witnessed by the
interference manner. We demonstrate the necessary and sufficient conditions
for the perfect two-photon interference. For a two-photon entangled state
with an anti-symmetric spectrum, it passes a 50/50 beam splitter with the
perfect transparency. The theory contributes an unified understanding to a
variety of the two-photon interference effects.
\end{abstract}

\begin{abstract}
PACS number(s): 42.50.Dv, 03.65.Ta, 42.50.Ar
\end{abstract}

\section{Introduction}

Photons as bosons tend to bunching, that is, photons are more likely to
appear close together than farther apart. This happens for the photonic
field with a classical analogy. For some optical field, however, photons may
behave as an opposite effect --- the photon anti-bunching. It is known that
photon anti-bunching effect provides an evidence for explicitly quantum
mechanical states of the optical field without classical analogy. When two
photons meet in a beamsplitter from two input ports, how do they exit from
the output ports? This is an interesting topic concerning the quantum nature
of photon interference. The first experimental observations of two-photon
interference in a beam splitter were reported in 1980s.\cite{mandel}\cite
{shih1} In the experiments, two photons to be interfered are spatially
separated and degenerated with the same frequency and polarization. This can
be done by the spontaneously parametric down-conversion (SPDC) of type I in
a crystal, in which a pair of photons, signal and idle, are produced. In the
degenerate case, two photons are mixed in a 50/50 beamsplitter, no
coincidence count of photons is found at two output ports. This effect was
lately called the photon coalescence interference (CI),\cite{gran} since two
photons meeting in beamsplitter go together. The early theoretical
explanation was based on indistinguishability of two single photons, that
is, the interference occurs for the degenerate photons when they meet in a
beamsplitter. In the further experiments, it has been found that, in
addition to the degenerate photons, the two-photon interference may occur
for two photons with different colors and polarizations.\cite{ou}-\cite{pa}
However, for the degenerate case when the individual signal photon and the
idle photon are arranged out of their coherent range (they do not meet at
the beamsplitter), the interference is still observed.\cite{chiao}-\cite{pit}
Obviously, these phenomena can not be explained by the indistinguishability
of two single photons. As a matter of fact, all the above experiments were
performed by the SPDC, the source emitting entangled photon pairs. A
successful theoretical explanation is to use two-photon entanglement with
the help of conceptual Feynman diagrams in which the pair of photons
interfered should be seen as a whole, the two-photon or biphoton, so that
the photon entanglement plays an essential role in two-photon interference.%
\cite{shih0}\cite{shih} Recently, the interference of two independent
photons has been studied experimentally and theoretically.\cite{san}-\cite
{by} Santori et al \cite{san} has demonstrated in their experiment, that two
independent single-photon pulses emitted by a semiconductor quantum dot show
a coalescence interference in a beamsplitter. In this case, it seems that
one cannot use the concept ''biphoton'' because the photon entanglement is
absent.

As we review all these studies on two-photon interference, it could be
confusing why sometimes two nondegenerate photons can interfere with each
other and sometimes they cannot, and sometimes the interference of two
degenerate photons occurs only when they meet together and sometimes the
photon meeting is not necessarily the case. A reasonable explanation could
be attributed to photon entanglement. As mentioned above, the entangled pair
for two nondegenerate photons should be seen as a biphoton and the
interference occurs between biphotons. Therefore, it might be concluded that
there are two kinds of interference mechanisms: the ''biphoton picture'' for
entangled photon pair and ''two photons picture'' for independent photons.
If that is true, one has to face a perplexed question, as argued in the
development of quantum mechanics, how the ''clever photons'' know whether
they should behave as a biphoton or a single-photon in the interference.

In this paper, we contribute a complete theoretical description for
two-photon interference in a beamsplitter. Since any realistic beam should
have a finite frequency range which must be taken into account, we study the
two-photon state in a wavepacket which can be either entangled or
un-entangled. The quantum descriptions of the beam splitter can be
implemented in both the Schr\"{o}dinger picture (S-picture) and the
Heisenberg picture (H-picture). Since the beamsplitter introduces a simple
transform for field operators, the theoretical description in the H-picture
is more convenient, and it was exploited in the most theoretical
discussions. Though the two pictures should give an identical result, the
description in the H-picture averages the distinct information on quantum
state. Instead, we discuss this issue in the S-picture, so it can show more
physical understanding for the nature of two-photon interference. The net
coincidence probability\ can be readily evaluated by our theory. We find
that the symmetry of two-photon spectrum plays a key role in the
interference manners. We distinguish the coalescence and anti-coalescence
interferences (ACI), and find out the necessary and sufficient condition for
the perfect CI and ACI. We prove that the photon entanglement is irrelevant
to CI, but necessary for ACI. Therefore, the ACI effect is the signature of
two-photon entanglement, and it could be an effective experimental method to
detect photon entanglement. However, we propose a two-photon transparent
state which can pass the beamsplitter with a full transparency. The theory
covers two cases: two-photon state in a polarization mode and in two
polarization modes, which may correspond to the source of SPDC of type I and
type II, respectively.

\section{Preliminary Theoretical Description}

\subsection{general description about quantum interference}

Let us briefly review how the quantum interference happens. If a quantum
system consists of more than one source, or the interaction includes several
parts, the state of the system $|\Psi \rangle $ is a coherent superposition
of these sources or parts 
\begin{eqnarray}
|\Psi _1\rangle &=&c_1|\alpha \rangle +c_2|\beta \rangle ,\qquad |\Psi
_2\rangle =c_3|\gamma \rangle +c_4|\delta \rangle ,  \label{b1} \\
|\Psi \rangle &=&|\Psi _1\rangle +|\Psi _2\rangle =c_1|\alpha \rangle
+c_2|\beta \rangle +c_3|\gamma \rangle +c_4|\delta \rangle .  \nonumber
\end{eqnarray}
Assume that all the states $|\alpha \rangle ,|\beta \rangle ,|\gamma \rangle 
$ and $|\delta \rangle $ are distinguishable each with others, there is no
quantum interference. If, however, there are some indistinguishable states
generated in these coherent sources such as 
\begin{eqnarray}
|\Psi _1\rangle &=&c_1|\alpha \rangle +c_2|\beta \rangle ,\qquad |\Psi
_2\rangle =c_3|\beta \rangle +c_4|\delta \rangle ,  \label{b2} \\
|\Psi \rangle &=&|\Psi _1\rangle +|\Psi _2\rangle =c_1|\alpha \rangle
+(c_2+c_3)|\beta \rangle +c_4|\delta \rangle ,  \nonumber
\end{eqnarray}
state $|\beta \rangle $ in two sources has to be added together. The
probability of finding state $|\beta \rangle $ for $|\Psi \rangle $ may not
be equal to the sum of those for $|\Psi _1\rangle $ and $|\Psi _2\rangle $,
unless the interference term 
\begin{equation}
c_2c_3^{*}+c_2^{*}c_3=2|c_2c_3|\cos [\arg (c_2)-\arg (c_3)]  \label{b3}
\end{equation}
is null. In other words, the quantum interference happens if the
interference term (\ref{b3}) is not null. There are two reasons for the
absence of interference. One possible reason is simply that there is no
indistinguishable state between two sources, that is $c_2=0$ or $c_3=0$.
Another reason could be out of phase for two coherent amplitudes, $c_2$ and $%
c_3$. In this sense, the indistinguishable state is only a necessary
condition for interference, but not a sufficient one. The relative phase
between the amplitudes $c_2$ and $c_3$ may settle the interference absence,
constructive and destructive according to null, positive and negative
interference terms, respectively.

In the language of quantum state, in essence, the interference originates
from the coherent superposition of probability amplitudes for the
indistinguishable states of different sources. We survey two-photon
interference in this picture.

\subsection{input-output transformation of quantum state in a beam splitter}

A beam splitter performs a linear transform for two input optical beams. In
the quantum regime, the bosonic commutation must be satisfied for all the
field operators in the beamsplitter transform. So, for a lossless
beamsplitter, the general transformation between input and output field
operators obeys\cite{cam} 
\begin{equation}
\left( 
\begin{array}{c}
b_1 \\ 
b_2
\end{array}
\right) =S(\theta ,\phi _\tau ,\phi _\rho )\left( 
\begin{array}{c}
a_1 \\ 
a_2
\end{array}
\right) ,\qquad S(\theta ,\phi _\tau ,\phi _\rho )=\left( 
\begin{array}{cc}
e^{i\phi _\tau }\cos \theta & e^{i\phi _\rho }\sin \theta \\ 
-e^{-i\phi _\rho }\sin \theta & e^{-i\phi _\tau }\cos \theta
\end{array}
\right) ,  \label{1}
\end{equation}
where $a_i$ and $b_i$ are the field annihilation operators for the input and
output ports, respectively. The subscript $i$ ($i=1,2$) symbolizes the ports
in the same propagation direction. $\theta $ characterizes the reflection
and the transmission rates, for instance, $\theta =\pi /4$ for a 50/50
beamsplitter. $\phi _\tau $ and $\phi _\rho $ are two phases allowed in the
unitary transformation (\ref{1}). In transformation (\ref{1}), two input
beams, $a_1$ and $a_2$, should be in the same mode. In other words, two
input photons are indistinguishable as soon as they are mixed in the
beamsplitter.

Transformation (\ref{1}) is carried out in the H-picture. In principle, it
can solve all the problems concerning the beamsplitter transform by
evaluating expectation values of field operators. Nevertheless, the method
wipes out the information of what happens for quantum state. This
shortcoming can be avoided in the S-picture. The transform of wavevector in
the S-picture, corresponding to the operator transformation (\ref{1}), $%
|\Psi \rangle _{out}=U|\Psi \rangle _{in}$, has been discussed in Ref. \cite
{cam}, in which an explicit expression of $U$ has been given. Alternatively,
we use a simpler method to evaluate the output wavevector. It is a similar
method as for a dynamic quantum system --- when the evolution of operators
has been known in the H-picture, one may obtain the state evolution in the
S-picture without solving the Schr\"{o}dinger equation.\cite{kaige} The
method requires two conditions: (i) the initial state is known as $|\Psi
\rangle _{in}=f(a_1,a_1^{\dagger },a_2,a_2^{\dagger })|\Theta \rangle $;
(ii) the inverse evolution of state $|\Theta \rangle $ is known as $|\Theta
^{\prime }\rangle =U^{-1}|\Theta \rangle $. In the beamsplitter case, we set 
$|\Theta \rangle =$ $|0\rangle $ and, obviously, a vacuum state $|0\rangle $
is always conserved as $U|0\rangle =U^{-1}|0\rangle =|0\rangle $. So that
the output state is obtained as 
\begin{eqnarray}
|\Psi \rangle _{out} &=&U|\Psi \rangle _{in}=Uf(a_1,a_1^{\dagger
},a_2,a_2^{\dagger })|0\rangle =Uf(a_1,a_1^{\dagger },a_2,a_2^{\dagger
})U^{-1}U|0\rangle =Uf(a_1,a_1^{\dagger },a_2,a_2^{\dagger })U^{-1}|0\rangle
\label{2} \\
&=&f(Ua_1U^{-1},Ua_1^{\dagger }U^{-1},Ua_2U^{-1},Ua_2^{\dagger
}U^{-1})|0\rangle =f(\overline{b}_1,\overline{b}_1^{\dagger },\overline{b}_2,%
\overline{b}_2^{\dagger })|0\rangle ,  \nonumber  \label{2}
\end{eqnarray}
where $\overline{b}_i\equiv Ua_iU^{-1}$ and $\overline{b}_i^{\dagger }\equiv
Ua_i^{\dagger }U^{-1}(i=1,2)$. Because $b_i=U^{-1}a_iU$ is known due to Eq. (%
\ref{1}), one may obtain its inverse transform as 
\begin{equation}
\left( 
\begin{array}{c}
\overline{b}_1 \\ 
\overline{b}_2
\end{array}
\right) =S^{-1}(\theta ,\phi _\tau ,\phi _\rho )\left( 
\begin{array}{c}
a_1 \\ 
a_2
\end{array}
\right) =S(-\theta ,-\phi _\tau ,\phi _\rho )\left( 
\begin{array}{c}
a_1 \\ 
a_2
\end{array}
\right) .  \label{3}
\end{equation}
It is not difficult to check $S(\theta ,\phi _\tau ,\phi _\rho )S(-\theta
,-\phi _\tau ,\phi _\rho )=I$. For the convenience in use, we write 
\begin{eqnarray}
\overline{b}_1^{\dagger }(\omega ) &=&\cos \theta e^{i\phi _\tau
}a_1^{\dagger }(\omega )-\sin \theta e^{-i\phi _\rho }a_2^{\dagger }(\omega
),  \label{4} \\
\overline{b}_2^{\dagger }(\omega ) &=&\sin \theta e^{i\phi _\rho
}a_1^{\dagger }(\omega )+\cos \theta e^{-i\phi _\tau }a_2^{\dagger }(\omega
).  \nonumber
\end{eqnarray}

\subsection{quantum states of single-photon and two-photon wavepackets}

A single-photon state of the monochromatic beam with the frequency $\omega $%
, travelling in a given propagation direction, is written as $a_\alpha
^{\dagger }(\omega )|0\rangle $. The index $\alpha $ denotes a particular
polarization or a spatial mode. Any practical beam has a finite bandwidth,
so a general form of the single photon state can be expressed as 
\begin{equation}
|\Phi _s\rangle =\sum_\omega C_\alpha (\omega )a_\alpha ^{\dagger }(\omega
)|0\rangle ,  \label{5}
\end{equation}
where $C_\alpha (\omega )$ is the spectrum of the probability amplitude. The
single-photon wavepacket corresponding to the above state is obtained as 
\begin{equation}
\langle 0|E_\alpha ^{(+)}(z,t)|\Phi _s\rangle =\sum_\omega {\cal E}_\alpha
(\omega )C_\alpha (\omega )e^{i\omega (z/c-t)}\equiv \widetilde{C}_\alpha
(z/c-t),  \label{6}
\end{equation}
where the field operator for the polarization mode $\alpha $ is described by 
\begin{equation}
E_\alpha ^{(+)}(z,t)\equiv \sum_\omega {\cal E}_\alpha (\omega )a_\alpha
(\omega )\exp [i\omega (z/c-t)],  \label{6p}
\end{equation}
and ${\cal E}_\alpha (\omega )$ is the field amplitude per photon. $|%
\widetilde{C}_\alpha (z/c-t)|^2$ shows an expectation for the field
intensity at $(z,t)$.

A single-photon state can be in a multimode superposition. The concept
''mode'' concerned here refers to polarization or spatial distribution, but
not radiation frequency and propagation direction, because the direction has
been given and the frequency dependence has been incorporated into state (%
\ref{5}). A single-photon state of two modes is written as 
\begin{equation}
|\Phi _s\rangle =\sum_\omega [C_\alpha (\omega )a_\alpha ^{\dagger }(\omega
)+C_\beta (\omega )a_\beta ^{\dagger }(\omega )]|0\rangle ,  \label{7}
\end{equation}
where $\alpha $ and $\beta $ are mode index.

Then we consider a quantum state containing two photons separated spatially.
Each photon has a given propagation direction, designated by the subscripts
1 and 2, so that the separated two photons are ready to be incident upon two
input ports of a beamsplitter. In the description of a two-photon
wavepacket, we discuss two cases.

$Case$ $I:$ two-photon wavepacket in the same polarization (or spatial) mode

If we assume two photons to be in the same polarization mode, the two-photon
state in a general form can be expressed as 
\begin{equation}
|\Phi _{two}\rangle =\sum_{\omega _1,\omega _2}C(\omega _1,\omega
_2)a_1^{\dagger }(\omega _1)a_2^{\dagger }(\omega _2)|0\rangle .  \label{10}
\end{equation}
Note that two photons in the state $a_1^{\dagger }(\omega _1)a_2^{\dagger
}(\omega _2)|0\rangle $ are distinguishable even if $\omega _1=\omega _2$
since they are separated physically. $C(\omega _1,\omega _2)$ denotes a
spectrum of two-photon wavepacket. The corresponding two-photon wavepacket
for state (\ref{10}) is given by 
\begin{equation}
\langle 0|E_1^{(+)}(z_1,t_1)E_2^{(+)}(z_2,t_2)|\Phi _{two}\rangle
=\sum_{\omega _1,\omega _2}{\cal E}(\omega _1){\cal E}(\omega _2)C(\omega
_1,\omega _2)e^{i\omega _1(z_1/c-t_1)}e^{i\omega _2(z_2/c-t_2)}\equiv 
\widetilde{C}(z_1/c-t_1,z_2/c-t_2).  \label{11}
\end{equation}
Equation (\ref{10}) can describe both entangled and un-entangled two-photon
states. If two-photon spectrum can be factorized as 
\begin{equation}
C(\omega _1,\omega _2)=C_1(\omega _1)C_2(\omega _2),  \label{8}
\end{equation}
the two-photon state $|\Phi _{two}\rangle $ is un-entangled. That is, the
two-photon state consists of two independent single-photon wavepackets such
as 
\begin{equation}
\langle 0|E_1^{(+)}(z_1,t_1)E_2^{(+)}(z_2,t_2)|\Phi _t\rangle =\sum_{\omega
_1,\omega _2}{\cal E}(\omega _1){\cal E}(\omega _2)C_1(\omega _1)C_2(\omega
_2)e^{i\omega _1(z_1/c-t_1)}e^{i\omega _2(z_2/c-t_2)}\equiv \widetilde{C}%
_1(z_1/c-t_1)\widetilde{C}_2(z_2/c-t_2),  \label{14}
\end{equation}
where $\widetilde{C}_j(z/c-t)$ is the single-photon wavepacket designated by
Eq. (\ref{6}). Otherwise, if the factorization (\ref{8}) is impossible, Eq. (%
\ref{10}) defines a frequency-entangled two-photon state. The corresponding
two-photon wavepacket (\ref{11}) can not be factorized into two
single-photon wavepackets as Eq. (\ref{14}) does.

This kind of two-photon state can be generated in the SPDC of type I. For
example, in the degenerate case, the two-photon spectrum of a pair of
entangled photons can be expressed in a symmetric form\cite{shih0} 
\begin{equation}
C(\omega _1,\omega _2)=g(\omega _1+\omega _2-\Omega _p)e^{-[(\omega
_1-\Omega )^2+(\omega _2-\Omega )^2]/(2\sigma ^2)},  \label{12}
\end{equation}
where $\Omega $ and $\sigma $ are the central frequency and the bandwidth,
respectively, for both the signal and the idle beams. $\Omega _p=2\Omega $
is the central frequency for the pump beam. Function $g(x)$ describes the
phase matching. For simplicity, it can be assumed as a Gaussian 
\begin{equation}
g(\omega _1+\omega _2-\Omega _p)=Ae^{-(\omega _1+\omega _2-\Omega
_p)^2/(2\sigma _p^2)},  \label{13}
\end{equation}
where $\sigma _p$ is the bandwidth of the pump beam. In the case of $\sigma
_p\rightarrow 0$, Eq. (\ref{13}) tends to a $\delta -$function 
\begin{equation}
g(\omega _1+\omega _2-\Omega _p)=A\delta (\omega _1+\omega _2-\Omega _p).
\label{13d}
\end{equation}
As an important example, a set of Bell states, which consist of two
monochromatic photons with frequencies $\Omega _1$ and $\Omega _2$ being in
the same polarization, are written as 
\begin{mathletters}
\label{bb}
\begin{eqnarray}
|\Phi ^{\pm }\rangle &=&(1/\sqrt{2})\sum_{\omega _1,\omega _2}[\delta
(\omega _1-\Omega _1)\delta (\omega _2-\Omega _1)\pm \delta (\omega
_1-\Omega _2)\delta (\omega _2-\Omega _2)]a_1^{\dagger }(\omega
_1)a_2^{\dagger }(\omega _2)|0\rangle ,  \label{b1a} \\
|\Psi ^{\pm }\rangle &=&(1/\sqrt{2})\sum_{\omega _1,\omega _2}[\delta
(\omega _1-\Omega _1)\delta (\omega _2-\Omega _2)\pm \delta (\omega
_1-\Omega _2)\delta (\omega _2-\Omega _1)]a_1^{\dagger }(\omega
_1)a_2^{\dagger }(\omega _2)|0\rangle .  \label{b1b}
\end{eqnarray}

$Case$ $II:$ two-photon wavepacket in two orthogonal polarization (or
spatial) modes

We assume that there are two un-entangled single-photon wavepackets
traveling in different directions, and each of them is described by a
two-mode superposition state (\ref{7}). The combined state for the two
photons is written as 
\end{mathletters}
\begin{eqnarray}
|\Phi _{ss}\rangle &=&\sum_{\omega _1}[C_{1\alpha }(\omega _1)a_{1\alpha
}^{\dagger }(\omega _1)+C_{1\beta }(\omega _1)a_{1\beta }^{\dagger }(\omega
_1)]\sum_{\omega _2}[C_{2\alpha }(\omega _2)a_{2\alpha }^{\dagger }(\omega
_2)+C_{2\beta }(\omega _2)a_{2\beta }^{\dagger }(\omega _2)]|0\rangle
\label{15} \\
&=&\sum_{\omega _1,\omega _2}[C_{1\alpha }(\omega _1)C_{2\alpha }(\omega
_2)a_{1\alpha }^{\dagger }(\omega _1)a_{2\alpha }^{\dagger }(\omega
_2)+C_{1\beta }(\omega _1)C_{2\beta }(\omega _2)a_{1\beta }^{\dagger
}(\omega _1)a_{2\beta }^{\dagger }(\omega _2)  \nonumber \\
&&+C_{1\alpha }(\omega _1)C_{2\beta }(\omega _2)a_{1\alpha }^{\dagger
}(\omega _1)a_{2\beta }^{\dagger }(\omega _2)+C_{1\beta }(\omega
_1)C_{2\alpha }(\omega _2)a_{1\beta }^{\dagger }(\omega _1)a_{2\alpha
}^{\dagger }(\omega _2)]|0\rangle .  \nonumber
\end{eqnarray}
In the general case, a two-mode two-photon wavepacket can be expressed as 
\begin{equation}
|\Phi _{two}\rangle =|\Phi _{\alpha \alpha }\rangle +|\Phi _{\beta \beta
}\rangle +|\Phi _{\alpha \beta }\rangle +|\Phi _{\beta \alpha }\rangle ,
\label{16}
\end{equation}
where 
\begin{mathletters}
\label{16p}
\begin{eqnarray}
|\Phi _{mm}\rangle &=&\sum_{\omega _1,\omega _2}C_{mm}(\omega _1,\omega
_2)a_{1m}^{\dagger }(\omega _1)a_{2m}^{\dagger }(\omega _2)|0\rangle ,\qquad
m=\alpha ,\beta  \label{16pa} \\
|\Phi _{\alpha \beta }\rangle &=&\sum_{\omega _1,\omega _2}C_{\alpha \beta
}(\omega _1,\omega _2)a_{1\alpha }^{\dagger }(\omega _1)a_{2\beta }^{\dagger
}(\omega _2)|0\rangle .\qquad \alpha \leftrightarrow \beta  \label{16pb}
\end{eqnarray}
There are four two-photon spectra, $C_{\alpha \alpha }(\omega _1,\omega
_2),C_{\beta \beta }(\omega _1,\omega _2),C_{\alpha \beta }(\omega _1,\omega
_2),C_{\beta \alpha }(\omega _1,\omega _2)$, which describe a two-mode
two-photon wavepacket. If the factorization of Eq. (\ref{16}) into Eq. (\ref
{15}) is impossible, the two-mode wavepacket is entangled.

This kind of two photon states can be generated in the SPDC of type II, in
which a pair of down-converted photons, o-ray and e-ray, are
polarization-orthogonal. A very famous example is the set of Bell states
consisting of two photons: one with frequency $\Omega _\alpha $ and
polarization $\alpha $ and the other one with frequency $\Omega _\beta $ and
polarization $\beta $, 
\end{mathletters}
\begin{mathletters}
\label{17}
\begin{eqnarray}
|\Phi ^{\pm }\rangle &=&(1/\sqrt{2})\sum_{\omega _1,\omega _2}[\delta
(\omega _1-\Omega _\alpha )\delta (\omega _2-\Omega _\alpha )a_{1\alpha
}^{\dagger }(\omega _1)a_{2\alpha }^{\dagger }(\omega _2)\pm \delta (\omega
_1-\Omega _\beta )\delta (\omega _2-\Omega _\beta )a_{1\beta }^{\dagger
}(\omega _1)a_{2\beta }^{\dagger }(\omega _2)]|0\rangle ,  \label{17a} \\
|\Psi ^{\pm }\rangle &=&(1/\sqrt{2})\sum_{\omega _1,\omega _2}[\delta
(\omega _1-\Omega _\alpha )\delta (\omega _2-\Omega _\beta )a_{1\alpha
}^{\dagger }(\omega _1)a_{2\beta }^{\dagger }(\omega _2)\pm \delta (\omega
_1-\Omega _\beta )\delta (\omega _2-\Omega _\alpha )a_{1\beta }^{\dagger
}(\omega _1)a_{2\alpha }^{\dagger }(\omega _2)]|0\rangle .  \label{17b}
\end{eqnarray}
The Bell states $|\Psi ^{\pm }\rangle $ can be generated in the SPDC process
of type-II, in which two down-converted photons come from the overlap of the
o-ray and e-ray cones.\cite{zei} However, a half-wave-plate can change the
polarization between horizontal and vertical, so that by using two
orthogonal half-wave-plates in two paths, the Bell states $|\Phi ^{\pm
}\rangle $ can be obtained from $|\Psi ^{\pm }\rangle $ (if one sets $\Omega
_\alpha =\Omega _\beta $). By taking into account the bandwidths of the
beams, the polarization entanglement state generated in SPDC of type II can
be described as 
\end{mathletters}
\begin{eqnarray}
|\Psi _w(\theta )\rangle &=&\sum_{\omega _1,\omega _2}g(\omega _1+\omega
_2-\Omega _p)[e^{-(\omega _1-\Omega _\alpha )^2/(2\sigma _\alpha ^2)-(\omega
_2-\Omega _\beta )^2/(2\sigma _\beta ^2)}a_{1\alpha }^{\dagger }(\omega
_1)a_{2\beta }^{\dagger }(\omega _2)  \label{18} \\
&&+e^{i\theta }e^{-(\omega _1-\Omega _\beta )^2/(2\sigma _\beta ^2)-(\omega
_2-\Omega _\alpha )^2/(2\sigma _\alpha ^2)}a_{1\beta }^{\dagger }(\omega
_1)a_{2\alpha }^{\dagger }(\omega _2)]|0\rangle ,  \nonumber
\end{eqnarray}
where $\Omega _p=\Omega _\alpha +\Omega _\beta ,$ and we have assumed that
o-ray and e-ray have different central frequencies and bandwidths. The phase 
$\theta $ depends on the crystal birefringence. But if one puts a wave-plate
in path 1, it is possible to introduce an additional relative phase to the
polarization $\beta $, so that the phase $\theta $ can be set as desired.%
\cite{zei}

\section{Two Photon Interference in a Beam Splitter}

\subsection{output quantum states and coincidence probability}

Equation (\ref{2}) can be used to calculate the output quantum state for any
input state incident upon a beamsplitter. We focus on the input states of a
two-photon wavepacket, as has been shown in the last section.

$Case$ $I:$ For the input state (\ref{10}), the corresponding output state
after the beamsplitter transform is obtained as 
\begin{eqnarray}
|\Phi _{two}\rangle _{out} &=&\sum_{\omega _1,\omega _2}C(\omega _1,\omega
_2)\overline{b}_1^{\dagger }(\omega _1)\overline{b}_2^{\dagger }(\omega
_2)|0\rangle  \label{19} \\
&=&\sum_{\omega _1,\omega _2}C(\omega _1,\omega _2)[a_1^{\dagger }(\omega
_1)e^{i\phi _\tau }\cos \theta -a_2^{\dagger }(\omega _1)e^{-i\phi _\rho
}\sin \theta ][a_1^{\dagger }(\omega _2)e^{i\phi _\rho }\sin \theta
+a_2^{\dagger }(\omega _2)e^{-i\phi _\tau }\cos \theta ]|0\rangle  \nonumber
\\
&=&\sum_{\omega _1,\omega _2}C(\omega _1,\omega _2)\{[a_1^{\dagger }(\omega
_1)a_1^{\dagger }(\omega _2)e^{i\phi }-a_2^{\dagger }(\omega _1)a_2^{\dagger
}(\omega _2)e^{-i\phi }]\cos \theta \sin \theta  \nonumber \\
&&+[a_1^{\dagger }(\omega _1)a_2^{\dagger }(\omega _2)\cos ^2\theta
-a_2^{\dagger }(\omega _1)a_1^{\dagger }(\omega _2)\sin ^2\theta
]\}|0\rangle ,  \nonumber
\end{eqnarray}
where $\phi =\phi _\tau +\phi _\rho $. In the summation taken in the whole
frequency space, the states corresponding to $(\omega _1,\omega _2)$ and $%
(\omega _2,\omega _1)$ are indistinguishable and should be added together.
For doing it, we may take $\sum_{\omega _1,\omega _2}=\sum_{\omega _1<\omega
_2}+\sum_{\omega _1=\omega _2}+\sum_{\omega _1>\omega _2}$, and then
exchange the variables $\omega _1$ and $\omega _2$ in the last summation. In
result, Eq. (\ref{19}) is written as 
\begin{eqnarray}
|\Phi _{two}\rangle _{out} &=&\sum_{\omega _1<\omega _2}\{[C(\omega
_1,\omega _2)+C(\omega _2,\omega _1)][a_1^{\dagger }(\omega _1)a_1^{\dagger
}(\omega _2)e^{i\phi }-a_2^{\dagger }(\omega _1)a_2^{\dagger }(\omega
_2)e^{-i\phi }]\cos \theta \sin \theta  \label{20} \\
&&+[C(\omega _1,\omega _2)\cos ^2\theta -C(\omega _2,\omega _1)\sin ^2\theta
]a_1^{\dagger }(\omega _1)a_2^{\dagger }(\omega _2)  \nonumber \\
&&+[C(\omega _2,\omega _1)\cos ^2\theta -C(\omega _1,\omega _2)\sin ^2\theta
]a_1^{\dagger }(\omega _2)a_2^{\dagger }(\omega _1)\}|0\rangle  \nonumber \\
&&+\sum_\omega C(\omega ,\omega )\{[(a_1^{\dagger }(\omega ))^2e^{i\phi
}-(a_2^{\dagger }(\omega ))^2e^{-i\phi }]\cos \theta \sin \theta +(\cos
^2\theta -\sin ^2\theta )a_1^{\dagger }(\omega )a_2^{\dagger }(\omega
)\}|0\rangle .  \nonumber
\end{eqnarray}
For a 50/50 beamsplitter (In the following text, we consider only the case
of the 50/50 beam splitter.), Eq. (\ref{20}) is reduced to 
\begin{eqnarray}
|\Phi _{two}\rangle _{out} &=&(1/2)\left( \sum_{\omega _1<\omega
_2}\{[C(\omega _1,\omega _2)+C(\omega _2,\omega _1)][a_1^{\dagger }(\omega
_1)a_1^{\dagger }(\omega _2)e^{i\phi }-a_2^{\dagger }(\omega _1)a_2^{\dagger
}(\omega _2)e^{-i\phi }]\right.  \label{21} \\
&&+[C(\omega _1,\omega _2)-C(\omega _2,\omega _1)][a_1^{\dagger }(\omega
_1)a_2^{\dagger }(\omega _2)-a_1^{\dagger }(\omega _2)a_2^{\dagger }(\omega
_1)]\}  \nonumber \\
&&\left. +\sum_\omega C(\omega ,\omega )[(a_1^{\dagger }(\omega ))^2e^{i\phi
}-(a_2^{\dagger }(\omega ))^2e^{-i\phi }]\right) |0\rangle .  \nonumber
\end{eqnarray}
In Eq. (\ref{21}), the first and last terms describe two photons exiting
from the same output port, whereas the second term describes two photons
exiting from two different ports resulting in a ''click-click'' in
coincidence measurement. The coincidence probability\ for a ''click-click''
detection at two output ports is evaluated by 
\begin{equation}
P_c=(1/4)\cdot 2\sum_{\omega _1<\omega _2}|C(\omega _1,\omega _2)-C(\omega
_2,\omega _1)|^2=(1/4)\sum_{\omega _1,\omega _2}|C(\omega _1,\omega
_2)-C(\omega _2,\omega _1)|^2,  \label{22}
\end{equation}
where we have considered that $|C(\omega _1,\omega _2)-C(\omega _2,\omega
_1)|^2$ is symmetric with respect to the diagonal $\omega _1=\omega _2$ in
the frequency space ($\omega _1,\omega _2$) and vanishes at $\omega
_1=\omega _2$. It can be expressed in the integration form as 
\begin{eqnarray}
P_c &=&(1/4)\int_{-\infty }^\infty d\omega _1\int_{-\infty }^\infty d\omega
_2|C(\omega _1,\omega _2)-C(\omega _2,\omega _1)|^2  \label{23} \\
&=&\frac 12\left( 1-\frac 12\int_{-\infty }^\infty d\omega _1\int_{-\infty
}^\infty d\omega _2[C(\omega _1,\omega _2)C^{*}(\omega _2,\omega _1)+\text{%
c.c.}]\right) ,  \nonumber
\end{eqnarray}
where the normalization $\int_{-\infty }^\infty d\omega _1\int_{-\infty
}^\infty d\omega _2|C(\omega _1,\omega _2)|^2=1$ has been applied.

What we have done from Eq. (\ref{19}) to Eq. (\ref{20}) is in fact the same
operation as Eq. (\ref{b2}). Physically, in the input two-photon wavepacket,
it may contain two sources:\\{\em source 1} -- a photon of frequency $\omega
_1$ at port 1 and the other photon of frequency $\omega _2$ at port 2 with
the amplitude $C(\omega _1,\omega _2)$;\\{\em source 2} -- a photon of
frequency $\omega _2$ at port 1 and the other photon of frequency $\omega _1$
at port 2 with the amplitude $C(\omega _2,\omega _1)$.\\The above pair of
sources, $a_1^{\dagger }(\omega _1)a_2^{\dagger }(\omega _2)|0\rangle $ and $%
a_1^{\dagger }(\omega _2)a_2^{\dagger }(\omega _1)|0\rangle $, in the input
state can generate indistinguishable output states. In the spectral plane
for the input state in which each point corresponds to a biphoton sub-state
with an amplitude $C(\omega _1,\omega _2)$, the diagonal $\omega _1=\omega
_2 $ divides the plane into two parts. The interference occurs between a
pair of symmetric points with respect to the diagonal, as shown in Fig. 1.

Equation (\ref{23}) for the coincidence probability shows an interference
term $C(\omega _1,\omega _2)C^{*}(\omega _2,\omega _1)+$c.c.. As pointed out
in Sec. II-A, the necessary and sufficient condition for the absence of
interference is 
\begin{equation}
\int_{-\infty }^\infty d\omega _1\int_{-\infty }^\infty d\omega _2[C(\omega
_1,\omega _2)C^{*}(\omega _2,\omega _1)+\text{c.c.}]=0.  \label{23a}
\end{equation}
It results in a 50\% coincidence probability, that is, the probability of
that two photons go together is equal to that they exit from different
ports. Obviously, if the spectrum of the input state satisfies 
\begin{equation}
|C(\omega _1,\omega _2)C(\omega _2,\omega _1)|=0,  \label{23c}
\end{equation}
the interference disappears. In this case, there is no pair of two-photon
states to be interfered, as shown in Fig. 1a. But condition (\ref{23c}) is
not necessary for the absence of interference. The other condition for the
absence of interference is out of phase between amplitudes $C(\omega
_1,\omega _2)$ and $C(\omega _2,\omega _1)$, i.e. $\arg C(\omega _1,\omega
_2)-\arg C(\omega _2,\omega _1)=(n+1/2)\pi $. Otherwise, the phase
difference of two biphoton states determines increase and decrease of the
coincidence probability.

Now we can answer the question raised in Introduction. In the language of
quantum state, it is clear to show the nature of interference based on the
''biphoton'', but not ''two photons''. What we emphasize is that this
interference mechanism does not ask for any precondition for the input
two-photon state, either entangled or un-entangled. We will show in the
following that entangled two-photon wavepacket behaves in a distinct
interference manner different from un-entangled one.

$Case$ $II:$ If we assume that the beamsplitter does not change the
polarization of the input beam, Eq. (\ref{4}) can be still used to the
polarization modes $\alpha $ and $\beta $ individually. In a two-photon
state with two polarizations, the biphoton state with the same polarization $%
\alpha \alpha $ is distinguishable from the states with the same
polarization $\beta \beta $ and the cross polarizations $\alpha \beta $ and $%
\beta \alpha $. Therefore, the interference can not occur among them. For
the input state shown in Eq. (\ref{16}), one may calculate the output state
in such a way 
\begin{mathletters}
\label{24}
\begin{eqnarray}
|\Phi _{two}\rangle _{out} &=&|\Phi _{\alpha \alpha }\rangle _{out}+|\Phi
_{\beta \beta }\rangle _{out}+|\Phi _{\alpha \beta +\beta \alpha }\rangle
_{out},  \label{24a} \\
|\Phi _{mm}\rangle _{out} &=&U|\Phi _{mm}\rangle ,\qquad m=\alpha ,\beta
\label{24b} \\
|\Phi _{\alpha \beta +\beta \alpha }\rangle _{out} &=&U(|\Phi _{\alpha \beta
}\rangle +|\Phi _{\beta \alpha }\rangle ).  \label{24c}
\end{eqnarray}
Equation (\ref{24b}) has already been calculated in Eqs. (\ref{19})-(\ref{21}%
). Equation (\ref{24c}) can be calculated as 
\end{mathletters}
\begin{eqnarray}
&&|\Phi _{\alpha \beta +\beta \alpha }\rangle _{out}=\sum_{\omega _1,\omega
_2}[C_{\alpha \beta }(\omega _1,\omega _2)\overline{b}_{1\alpha }^{\dagger
}(\omega _1)\overline{b}_{2\beta }^{\dagger }(\omega _2)+C_{\beta \alpha
}(\omega _1,\omega _2)\overline{b}_{1\beta }^{\dagger }(\omega _1)\overline{b%
}_{2\alpha }^{\dagger }(\omega _2)]|0\rangle  \label{25} \\
&=&\sum_{\omega _1,\omega _2}[C_{\alpha \beta }(\omega _1,\omega _2)%
\overline{b}_{1\alpha }^{\dagger }(\omega _1)\overline{b}_{2\beta }^{\dagger
}(\omega _2)+C_{\beta \alpha }(\omega _2,\omega _1)\overline{b}_{1\beta
}^{\dagger }(\omega _2)\overline{b}_{2\alpha }^{\dagger }(\omega
_1)]|0\rangle  \nonumber \\
&=&\sum_{\omega _1,\omega _2}\{[C_{\alpha \beta }(\omega _1,\omega
_2)+C_{\beta \alpha }(\omega _2,\omega _1)][a_{1\alpha }^{\dagger }(\omega
_1)a_{1\beta }^{\dagger }(\omega _2)e^{i\phi }-a_{2\alpha }^{\dagger
}(\omega _1)a_{2\beta }^{\dagger }(\omega _2)e^{-i\phi }]\sin \theta \cos
\theta  \nonumber \\
&&+[C_{\alpha \beta }(\omega _1,\omega _2)\cos ^2\theta -C_{\beta \alpha
}(\omega _2,\omega _1)\sin ^2\theta ]a_{1\alpha }^{\dagger }(\omega
_1)a_{2\beta }^{\dagger }(\omega _2)  \nonumber \\
&&-[C_{\alpha \beta }(\omega _1,\omega _2)\sin ^2\theta -C_{\beta \alpha
}(\omega _2,\omega _1)\cos ^2\theta ]a_{2\alpha }^{\dagger }(\omega
_1)a_{1\beta }^{\dagger }(\omega _2)\}|0\rangle .  \nonumber
\end{eqnarray}
For a 50/50 beamsplitter, it is written as 
\begin{eqnarray}
|\Phi _{\alpha \beta +\beta \alpha }\rangle _{out} &=&(1/2)\sum_{\omega
_1,\omega _2}\{[C_{\alpha \beta }(\omega _1,\omega _2)+C_{\beta \alpha
}(\omega _2,\omega _1)][a_{1\alpha }^{\dagger }(\omega _1)a_{1\beta
}^{\dagger }(\omega _2)e^{i\phi }-a_{2\alpha }^{\dagger }(\omega
_1)a_{2\beta }^{\dagger }(\omega _2)e^{-i\phi }]  \label{26} \\
&&+[C_{\alpha \beta }(\omega _1,\omega _2)-C_{\beta \alpha }(\omega
_2,\omega _1)][a_{1\alpha }^{\dagger }(\omega _1)a_{2\beta }^{\dagger
}(\omega _2)-a_{2\alpha }^{\dagger }(\omega _1)a_{1\beta }^{\dagger }(\omega
_2)]\}|0\rangle .  \nonumber
\end{eqnarray}
In the summation, the first term shows $\alpha $ and $\beta $ photons
traveling together, and the second term shows $\alpha $ and $\beta $ photons
exiting separately from two output ports, causing a ''click-click'' counting.

For the input state of Eq. (\ref{16}), the normalization is described as 
\begin{mathletters}
\label{27}
\begin{eqnarray}
1 &=&n_{\alpha \alpha }+n_{\beta \beta }+n_{\alpha \beta }+n_{\beta \alpha },
\label{27a} \\
n_{mm} &=&\int_{-\infty }^\infty d\omega _1\int_{-\infty }^\infty d\omega
_2|C_{mm}(\omega _1,\omega _2)|^2,\qquad m=\alpha ,\beta ,  \label{27b} \\
n_{\alpha \beta } &=&\int_{-\infty }^\infty d\omega _1\int_{-\infty }^\infty
d\omega _2|C_{\alpha \beta }(\omega _1,\omega _2)|^2,\qquad \alpha
\leftrightarrow \beta ,  \label{27c}
\end{eqnarray}
where $n_{ij}$ indicates the probability proportion of the input state $%
|\Phi _{ij}\rangle $ of Eq. (\ref{16}).

For case II, we consider polarization-sensitivity of detection system which
can distinguish the output coincidence probability for two photons with a
particular configuration of polarizations. Similar to Eq. (\ref{23}), the
coincidence probability\ for the same polarized photons is obtained as 
\end{mathletters}
\begin{eqnarray}
P_c^{mm} &=&(1/4)\int_{-\infty }^\infty d\omega _1\int_{-\infty }^\infty
d\omega _2|C_{mm}(\omega _1,\omega _2)-C_{mm}(\omega _2,\omega _1)|^2\qquad
(m=\alpha ,\beta )  \label{28} \\
&=&\frac 12n_{mm}\left( 1-\frac 1{2n_{mm}}\int_{-\infty }^\infty d\omega
_1\int_{-\infty }^\infty d\omega _2[C_{mm}(\omega _1,\omega
_2)C_{mm}^{*}(\omega _2,\omega _1)+\text{c.c.}]\right) ,  \nonumber
\end{eqnarray}
where the normalization (\ref{27b}) has been used. The coincidence
probabilities for two output photons with the cross polarizations can be
obtained by Eq. (\ref{26}). Two coincidence probabilities, $\alpha (\beta )$
photon at port 1 and $\beta (\alpha )$ photon at port 2, are the same as 
\begin{eqnarray}
P_c^{\alpha \beta } &=&P_c^{\beta \alpha }=(1/4)\int_{-\infty }^\infty
d\omega _1\int_{-\infty }^\infty d\omega _2|C_{\alpha \beta }(\omega
_1,\omega _2)-C_{\beta \alpha }(\omega _2,\omega _1)|^2  \label{29} \\
&=&\frac 14(n_{\alpha \beta }+n_{\beta \alpha })\left( 1-\frac 1{n_{\alpha
\beta }+n_{\beta \alpha }}\int_{-\infty }^\infty d\omega _1\int_{-\infty
}^\infty d\omega _2[C_{\alpha \beta }(\omega _1,\omega _2)C_{\beta \alpha
}^{*}(\omega _2,\omega _1)+\text{c.c.}]\right) ,  \nonumber
\end{eqnarray}
where the normalization (\ref{27c}) has been used. If the detection does not
distinguish polarization, the total coincidence probability is detected as 
\begin{eqnarray}
P_c &=&P_c^{\alpha \alpha }+P_c^{\beta \beta }+2P_c^{\alpha \beta }
\label{29t} \\
&=&\frac 12\left\{ 1-\frac 12\int_{-\infty }^\infty d\omega _1\int_{-\infty
}^\infty d\omega _2[2C_{\alpha \beta }(\omega _1,\omega _2)C_{\beta \alpha
}^{*}(\omega _2,\omega _1)+\sum_{m=\alpha ,\beta }C_{mm}(\omega _1,\omega
_2)C_{mm}^{*}(\omega _2,\omega _1)+\text{c.c.}]\right\} .  \nonumber
\end{eqnarray}

As mentioned above, because of the distinguishability of the polarization
configuration, the interference between two photons with the same
polarization is independent of that for the cross polarizations. The
condition of the absence of interference for the same polarized photons is
the same as that in case I. (see Eq. (\ref{23a})) As for the states $|\Phi
_{\alpha \beta }\rangle $ and $|\Phi _{\beta \alpha }\rangle $ shown in Eq. (%
\ref{16pb}), the condition for the absence of two-photon interference is a
null interference term 
\begin{equation}
\int_{-\infty }^\infty d\omega _1\int_{-\infty }^\infty d\omega _2[C_{\alpha
\beta }(\omega _1,\omega _2)C_{\beta \alpha }^{*}(\omega _2,\omega _1)+\text{%
c.c.}]=0.  \label{27p}
\end{equation}
It results in the coincidence probability of the cross-polarized photons 
\begin{equation}
P_c^{\alpha \beta }=P_c^{\beta \alpha }=\frac 14(n_{\alpha \beta }+n_{\beta
\alpha }),  \label{27cp}
\end{equation}
so that the coincidence probability $P_c^{\alpha \beta }+P_c^{\beta \alpha
}=2P_c^{\alpha \beta }$ is one half of the probability proportion $n_{\alpha
\beta }+n_{\beta \alpha }$ for the pairs of cross-polarized photons. Again,
the probability that two photons go together is the same as that they exit
separately. Equation (\ref{29}) shows that the interference occurs between
the input states $|\Phi _{\alpha \beta }\rangle $ and $|\Phi _{\beta \alpha
}\rangle $. If only one state, either $|\Phi _{\alpha \beta }\rangle $ or $%
|\Phi _{\beta \alpha }\rangle $, exists, the interference never happens
because of 
\begin{equation}
|C_{\alpha \beta }(\omega _1,\omega _2)C_{\beta \alpha }(\omega _2,\omega
_1)|=0.  \label{27sc}
\end{equation}
Similarly, the phase difference between two amplitudes, $C_{\alpha \beta
}(\omega _1,\omega _2)$ and $C_{\beta \alpha }(\omega _2,\omega _1)$,
dominates the occurrence of interference.

When both the input states $|\Phi _{\alpha \beta }\rangle $ and $|\Phi
_{\beta \alpha }\rangle $ coexist, the interference occurs between two
sources:\\{\em source 1} -- a photon $(\omega _1,\alpha )$ at port 1 and the
other photon $(\omega _2,\beta )$ at port 2 with the amplitude $C_{\alpha
\beta }(\omega _1,\omega _2)$;\\{\em source 2} -- a photon $(\omega _2,\beta
)$ at port 1 and the other photon $(\omega _1,\alpha )$ at port 2 with the
amplitude $C_{\beta \alpha }(\omega _2,\omega _1).$\\Sources 1 and 2 come
from the input states $|\Phi _{\alpha \beta }\rangle $ and $|\Phi _{\beta
\alpha }\rangle $, respectively. Note that in this case two photons at two
input ports are always orthogonal in polarization and there is no degenerate
photons. Undoubtedly, the effect cannot be understood in the ''two photons
picture''. To understand the interference mechanism in case II, we give a
simple explanation. The pair of sources, $a_{1\alpha }^{\dagger }(\omega
_1)a_{2\beta }^{\dagger }(\omega _2)|0\rangle $ and $a_{1\beta }^{\dagger
}(\omega _2)a_{2\alpha }^{\dagger }(\omega _1)|0\rangle ,$ become
indistinguishable when they are mixed in the beamsplitter. This can be seen
by omitting the subscripts 1 and 2 of the creating operators. However, the
pair of sources, $a_{1\alpha }^{\dagger }(\omega _1)a_{2\beta }^{\dagger
}(\omega _2)|0\rangle $ and $a_{1\alpha }^{\dagger }(\omega _2)a_{2\beta
}^{\dagger }(\omega _1)|0\rangle ,$ are still distinguishable as the
subscripts 1 and 2 have been omitted. So that the interference can not occur
when the state $|\Phi _{\alpha \beta }\rangle $ (or $|\Phi _{\beta \alpha
}\rangle $) exists by itself even if it has a symmetric spectrum $C_{\alpha
\beta }(\omega _1,\omega _2)=C_{\alpha \beta }(\omega _2,\omega _1) $.

\subsection{coalescence interference}

In the last subsection, we show the interference term in the representation
of quantum state. In the presence of interference, the interference term
increases or decreases the coincidence probability with respect to that of
the absence of interference. It is necessary to define two manners of
two-photon interferences, the coalescence interference (CI) and the
anti-coalescence interferences (ACI), according to the coincidence
probabilities less and more than that for the absence of interference,
respectively. For the CI effect, the probability of the fact that two
photons travel together is more than the probability of the fact that they
exit apart. In the extreme case, two photons always go together and the
coincidence probability is null, one calls it the perfect coalescence
interference.

$Case$ $I:$ According to Eqs. (\ref{21}) or (\ref{23}), the sufficient and
necessary condition for the perfect CI is that the two-photon wavepacket has
a symmetric spectrum in the whole frequency range 
\begin{equation}
C(\omega _1,\omega _2)\equiv C(\omega _2,\omega _1).  \label{ss}
\end{equation}
A symmetric spectrum can be acquired for both entangled and un-entangled
two-photon wavepackets. For example, a pair of degenerate photons generated
in SPDC of type I, such as shown in Eq. (\ref{12}), has a symmetric
spectrum. The three Bell states $|\Phi ^{\pm }\rangle $ and $|\Psi
^{+}\rangle $ described in Eqs. (\ref{bb}) are also symmetric. These
examples of two-photon entangled states show the perfect CI.

For two independent single-photon wavepackets, the two-photon spectrum is
the product of two single-photon spectra shown in Eq. (\ref{8}). If two
single-photon spectra are identical, $C_1(\omega )=C_2(\omega )=C(\omega )$,
the symmetric condition (\ref{ss}) is fulfilled. This means two identical
single-photon wavepackets perform the perfect CI. On the contrary, if the
symmetric spectrum (\ref{ss}) is satisfied for two independent wavepackets,
i.e. $C_1(\omega _1)C_2(\omega _2)\equiv C_2(\omega _1)C_1(\omega _2)$, it
has 
\begin{equation}
\frac{C_1(\omega _1)}{C_2(\omega _1)}\equiv \frac{C_1(\omega _2)}{C_2(\omega
_2)}=s,  \label{29s}
\end{equation}
where $s$ is a constant independent of frequency. By taking into account the
normalization $1=\int |C_1(\omega )|^2d\omega =\int |C_2(\omega )|^2d\omega $%
, one obtains $s=e^{i\theta }$ and hence 
\begin{equation}
C_1(\omega )=e^{i\theta }C_2(\omega ).  \label{29p}
\end{equation}
The spectrum of two un-entangled single-photon wavepackets is then $C(\omega
_1,\omega _2)=e^{i\theta }C_2(\omega _1)C_2(\omega _2)$. The phase factor
independent of frequency is actually trivial. In this sense, we can conclude
that the necessary and sufficient condition for the perfect CI of two
independent single-photon wavepackets is that two single-photon wavepackets
are identical.

Furthermore, we prove that, for two independent single-photon wavepackets,
the coincidence probability is not greater than one half. Since one has 
\begin{eqnarray}
&&\int_{-\infty }^\infty d\omega _1\int_{-\infty }^\infty d\omega _2C(\omega
_1,\omega _2)C^{*}(\omega _2,\omega _1)  \label{30} \\
&=&\int_{-\infty }^\infty d\omega _1\int_{-\infty }^\infty d\omega
_2C_1(\omega _1)C_2(\omega _2)C_1^{*}(\omega _2)C_2^{*}(\omega
_1)=|\int_{-\infty }^\infty d\omega C_1(\omega )C_2^{*}(\omega )|^2, 
\nonumber
\end{eqnarray}
Eq. (\ref{23}) is written as 
\begin{equation}
P_c=\frac 12\left[ 1-\left| \int_{-\infty }^\infty d\omega C_1(\omega
)C_2^{*}(\omega )\right| ^2\right] \leq \frac 12.  \label{30p}
\end{equation}
This is the right reason why we distinguish CI and ACI effects. In case I,
the ACI effect\ never happens for two un-entangled single-photon wavepackets.

Equation (\ref{30p}) shows that if two independent single-photon spectra
never overlap in the whole frequency range, i.e. $|C_1(\omega )C_2(\omega
)|\equiv 0$, there is no two-photon interference. In the ''two photons
picture'', it would be explained by the distinguishability of two input
photons, i.e. two photons have different frequencies. But this explanation
is inconsistent with the fact that two non-degenerate photons can interfere
in some other cases. However, in the ''biphoton picture'', we can find the
correct understanding. To be explicit, we assume that the spectrum of each
single-photon wavepacket is a Gaussian type $C_i(\omega )\sim \exp [-(\omega
-\Omega _i)^2/(2\sigma ^2)]$ ($i=1,2$). If the two single-photon spectra
have the same central frequency $\Omega _1=\Omega _2=\Omega $, i.e. they are
identical, the corresponding two-photon spectrum $C(\omega _1,\omega _2)\sim
\exp [-((\omega _1-\Omega )^2+(\omega _2-\Omega )^2)/(2\sigma ^2)]$ is
symmetric with respect to the diagonal $\omega _1=\omega _2$, as shown by
the contour plot of the spectrum in Fig. 2a. If, however, the difference of
the two central frequencies is larger than the bandwidth, $|\Omega _2-\Omega
_1|>\sigma $, the two single-photon spectra do not overlap. In the spectral
space for the two-photon states, the centre of the two-photon spectrum
deviates from the diagonal, as shown in Fig. 2b. Similar to Fig. 1a, there
are few pairs of photons to be interfered.

In this theory, the net coincidence probability can be calculated for
showing the manners of two-photon interference. Experimentally, it would be
difficult to detect the net coincidence probability because of the lower
quantum detection efficiency. A simple way is to compare the relative value
of the coincidence probability with respect to a reference, for example, the
one for the absence of interference. In experiment, this can be done by
introducing different paths for two incident beams. Let us assume a
two-photon spectrum $C_s(\omega _1,\omega _2)$ to be defined at an optical
source emitting two beams separately. These two beams, traveling different
paths $z_1$ and $z_2,$ are ready to input into two ports of a beamsplitter.
According to Eq. (\ref{11}), the new spectrum after the paths is written as 
\begin{equation}
C(\omega _1,\omega _2)=C_s(\omega _1,\omega _2)e^{i(\omega _1z_1/c+\omega
_2z_2/c)}.  \label{31}
\end{equation}
If $C_s(\omega _1,\omega _2)$ is symmetric, $C(\omega _1,\omega _2)$ becomes
asymmetric at the unbalanced position $z_1\neq z_2$.

As an example, we consider a source emitting a two-photon wavepacket with
the spectrum as 
\begin{equation}
C_s(\omega _1,\omega _2)=g(\omega _1+\omega _2-\Omega _p)e^{-[(\omega
_1-\Omega _1)^2+(\omega _2-\Omega _2)^2]/(2\sigma ^2)},  \label{31n}
\end{equation}
where $\Omega _p=\Omega _1+\Omega _2$ describes the phase matching in which $%
\Omega _p$ and $\Omega _i$ ($i=1,2$) are the central frequencies for the
pump beam and two converted beams, respectively. $\sigma $ defines the
bandwidth for two converted beams. This is the typical form of two-photon
wavepacket generated in SPDC of type I by taking into account the two
down-converted beams non-degenerate. If $g(\omega _1+\omega _2-\Omega _p)$
can not be factorized as $g_1(\omega _1)g_2(\omega _2)$, it describes an
entangled two-photon wavepacket. However, Eq. (\ref{31n}) can describe an
un-entangled two-photon wavepacket too, as long as $g(x)=1$. For the
spectrum (\ref{31n}), we calculate the coincidence probability by Eq. (\ref
{23}) and obtain (see Appendix A) 
\begin{equation}
P_c=(1/2)[1-e^{-\frac 12\Delta z^2(\sigma /c)^2}e^{-\frac 12(\Delta \Omega
/\sigma )^2}],  \label{32}
\end{equation}
where the path difference $\Delta z=z_2-z_1$ and the frequency deference $%
\Delta \Omega =\Omega _2-\Omega _1$. The equation displays a well known
interference dip at the balanced position $z_1=z_2$, observed in the
previous experiments, for example, in Refs.\cite{mandel} and \cite{shih1}
for the entangled two-photon state, and in the recent experiment reported in
Ref. \cite{san} for two independent single photons. The width of the dip is
defined by the coherent length of the single-photon beam $c/\sigma $. When
the path difference $\Delta z$ exceeds far the coherent length $\Delta
z>>c/\sigma ,$ the CI disappears, showing the reference ($P_c=1/2$) for the
absence of interference. In the degenerate case $\Delta \Omega =0$, the
coincidence probability is null at the balanced position and the perfect CI
occurs due to the symmetry of the spectrum (\ref{31n}). The level of dip
rises as the difference of the central frequencies $\Delta \Omega $ is
increased. However, it is interesting that the coincidence probability does
not depend on the form of function $g(x)$, so that the present theory
contributes an uniform description for both entangled and un-entangled
two-photon wavepackets.

In a general case when the bandwidths of two single-photon beams are not
equal, in order to evaluate the coincidence probability, function $g(x)$
must be given. Let the spectrum of two-photon wavepacket 
\begin{equation}
C_s(\omega _1,\omega _2)=Ae^{-(\omega _1+\omega _2-\Omega _p)^2/(2\sigma
_p^2)}e^{-(\omega _1-\Omega _1)^2/(2\sigma _1^2)-(\omega _2-\Omega
_2)^2/(2\sigma _2^2)},  \label{32na}
\end{equation}
where $\sigma _p$ is the bandwidth for the pump beam. The coincidence
probability is calculated as 
\begin{equation}
P_c=(1/2)[1-(\sigma _s/\sigma _f)e^{-\frac 12\Delta z^2(\sigma _s/c)^2}e^{-%
\frac 12(\Delta \Omega /\sigma _f)^2}],  \label{32pc}
\end{equation}
where two effective bandwidths are defined as 
\begin{mathletters}
\label{32s}
\begin{eqnarray}
\sigma _s &=&\sqrt{\frac{2\sigma _1^2\sigma _2^2}{\sigma _1^2+\sigma _2^2}},
\label{32sa} \\
\sigma _f &=&\sqrt{\frac{\sigma _p^2(\sigma _1^2+\sigma _2^2)+4\sigma
_1^2\sigma _2^2}{2(\sigma _p^2+\sigma _1^2+\sigma _2^2)}}.  \label{32sb}
\end{eqnarray}
The effective bandwidths $\sigma _s$ and $\sigma _f$ determine the spatial
coherent range and the frequency range of two-photon interference,
respectively. In the extreme case $\sigma _p\rightarrow 0$ which describes
the maximum two-photon entanglement, two effective bandwidths are equal: $%
\sigma _f=\sigma _s$. However, when $\sigma _p\rightarrow \infty $, the two
converted beams are not entangled, the coincidence probability is written as 
\end{mathletters}
\begin{equation}
P_c=\frac 12\left[ 1-\frac{2\sigma _1\sigma _2}{\sigma _1^2+\sigma _2^2}\exp
(-\frac{\sigma _1^2\sigma _2^2}{\sigma _1^2+\sigma _2^2}\cdot \frac{\Delta
z^2}{c^2}-\frac{\Delta \Omega ^2}{\sigma _1^2+\sigma _2^2})\right] .
\label{32id}
\end{equation}
It verifies $P_c<1/2$ for two independent single-photon wavepackets.

$Case$ $II:$ In the general form of case II, the input state includes four
parts shown in Eq. (\ref{16}). We have already indicated in Sec. III-A that
there is no interference among the states $|\Phi _{\alpha \alpha }\rangle
,|\Phi _{\beta \beta }\rangle $ and $|\Phi _{\alpha \beta }\rangle +|\Phi
_{\beta \alpha }\rangle $, so that they can be discussed independently. For
state $|\Phi _{mm}\rangle $, discussion is the same as case I. For $|\Phi
_{\alpha \beta }\rangle +|\Phi _{\beta \alpha }\rangle $, one may use Eq. (%
\ref{26}), or Eq. (\ref{29}) to study the CI effect. Therefore, for case II,
the sufficient and necessary conditions for the perfect CI are obtained as 
\begin{mathletters}
\label{33}
\begin{eqnarray}
C_{mm}(\omega _1,\omega _2) &\equiv &C_{mm}(\omega _2,\omega _1),\qquad
m=\alpha ,\beta  \label{33a} \\
C_{\alpha \beta }(\omega _1,\omega _2) &\equiv &C_{\beta \alpha }(\omega
_2,\omega _1).  \label{33b}
\end{eqnarray}
We note that the conditions (\ref{33a}) and (\ref{33b}) are for the perfect
CI of two photons with the same polarization (state $|\Phi _{mm}\rangle $)
and with orthogonal polarizations (state $|\Phi _{\alpha \beta }\rangle
+|\Phi _{\beta \alpha }\rangle $), respectively.

Similar to case I, the perfect CI can be acquired for both entangled and
un-entangled two-photon wavepacket. For the entangled two-photon states, for
instance, the Bell states $|\Phi ^{\pm }\rangle $ and $|\Psi ^{+}\rangle $
described by Eqs. (\ref{17a}) and (\ref{17b}) satisfy the symmetric
conditions (\ref{33a}) and (\ref{33b}), respectively. However, the
polarization-entangled two-photon wavepacket $|\Psi _w(\theta =0)\rangle ,$
defined by Eq. (\ref{18}), satisfies condition (\ref{33b}).

Then we consider two independent single-photon wavepackets, described by Eq.
(\ref{15}). The normalization of each single-photon wavepacket requires 
\end{mathletters}
\begin{equation}
n_{jm}=\int d\omega |C_{jm}(\omega )|^2,\qquad n_{j\alpha }+n_{j\beta
}=1,\qquad (j=1,2,\quad m=\alpha ,\beta ).  \label{sn}
\end{equation}
If two single-photon wavepackets are identical 
\begin{equation}
C_{1m}(\omega )=C_{2m}(\omega ),\qquad m=\alpha ,\beta  \label{40}
\end{equation}
it is readily to verify that the symmetric conditions (\ref{33}) have been
satisfied. As the same as case I, two identical single-photon wavepackets
show the perfect CI. On the other hand, if the symmetric condition Eq. (\ref
{33b}) has been satisfied for two independent single-photon wavepackets, one
obtains 
\begin{equation}
\frac{C_{1\alpha }(\omega _1)}{C_{2\alpha }(\omega _1)}=\frac{C_{1\beta
}(\omega _2)}{C_{2\beta }(\omega _2)}=s,  \label{37}
\end{equation}
where the constant $s$ is independent of frequency. By taking into account
the normalization (\ref{sn}), one obtains $s=e^{i\theta }$, and hence 
\begin{equation}
C_{1\alpha }(\omega )=e^{i\theta }C_{2\alpha }(\omega ),\qquad C_{1\beta
}(\omega )=e^{i\theta }C_{2\beta }(\omega ).  \label{39}
\end{equation}
This means that, for two independent single-photon wavepackets, if only
condition (\ref{33b}) has been satisfied, condition (\ref{33a}) must be
satisfied, too, and the two wavepackets are identical in addition to a
phase. In other words, for two independent single-photon wavepackets, if the
perfect CI for photons with the cross polarizations has been observed, one
can predict the perfect CI for photons with the same polarization.

We calculate the coincidence probability for two independent single-photon
wavepackets. Using Eqs. (\ref{28}) and (\ref{30}), we obtain 
\begin{equation}
P_c^{mm}=\frac 12[n_{1m}n_{2m}-|\int d\omega C_{1m}(\omega
)C_{2m}^{*}(\omega )|^2],\qquad m=\alpha ,\beta ,  \label{42}
\end{equation}
where $n_{1m}n_{2m}$ is the probability of two $m$-polarized photons
entering the beamsplitter. Equation (\ref{42}) shows that $(1/2)n_{1m}n_{2m}$
is the reference coincidence probability for the absence of interference of
the $m$-polarized photons. This means that the ACI effect cannot be observed
in detection of the coincidence probability of the same polarized photons.
By taking into account the integral 
\begin{eqnarray}
\int_{-\infty }^\infty d\omega _1\int_{-\infty }^\infty d\omega _2C_{\alpha
\beta }(\omega _1,\omega _2)C_{\beta \alpha }^{*}(\omega _2,\omega _1)
&=&\int_{-\infty }^\infty d\omega _1\int_{-\infty }^\infty d\omega
_2C_{1\alpha }(\omega _1)C_{2\beta }(\omega _2)C_{1\beta }^{*}(\omega
_2)C_{2\alpha }^{*}(\omega _1)  \label{34} \\
&=&\int_{-\infty }^\infty d\omega C_{1\alpha }(\omega )C_{2\alpha
}^{*}(\omega )\times \int_{-\infty }^\infty d\omega C_{2\beta }(\omega
)C_{1\beta }^{*}(\omega ),  \nonumber
\end{eqnarray}
and Eq. (\ref{29}), the coincidence probability for the pairs of
cross-polarized photons is obtained as 
\begin{equation}
P_c^{\alpha \beta }=P_c^{\beta \alpha }=\frac 14(n_{1\alpha }n_{2\beta
}+n_{1\beta }n_{2\alpha })-\frac 14[\int_{-\infty }^\infty d\omega
C_{1\alpha }(\omega )C_{2\alpha }^{*}(\omega )\times \int_{-\infty }^\infty
d\omega C_{2\beta }(\omega )C_{1\beta }^{*}(\omega )+\text{c.c.}],
\label{41}
\end{equation}
where $n_{1\alpha }n_{2\beta }+n_{1\beta }n_{2\alpha }$ is the probability
of two cross-polarized photons entering the beamsplitter. Again, $%
(1/2)(n_{1\alpha }n_{2\beta }+n_{1\beta }n_{2\alpha })$ is the reference
coincidence probability for the absence of interference for two
cross-polarized photons. Different from $P_c^{mm}$, $P_c^{\alpha \beta }+$ $%
P_c^{\beta \alpha }$can be higher or lower than this reference. Finally, the
total coincidence probability (\ref{29t}) is written as 
\begin{equation}
P_c=\frac 12\left[ 1-\left| \int_{-\infty }^\infty d\omega [C_{1\alpha
}(\omega )C_{2\alpha }^{*}(\omega )+C_{1\beta }(\omega )C_{2\beta
}^{*}(\omega )]\right| ^2\right] \leq \frac 12,  \label{35}
\end{equation}
where the normalization (\ref{sn}) has been applied. This result tells us
that, for two independent single-photon wavepackets, the ACI effect never
occurs for the polarization-insensitive detection. Of course, each
sub-coincidence probability, such as $P_c^{mm}(m=\alpha ,\beta )$ or $%
P_c^{\alpha \beta }+$ $P_c^{\beta \alpha }$, is no more than one half, too.

In consequence, two-photon coalescence interference in a beamsplitter can
inspect the identity for two independent single-photon wavepackets or,
substantially, the symmetry of the spectrum of two-photon wavepacket.
Obviously, CI is not a criterion for two-photon entanglement. We have proved
in Eqs. (\ref{30p}) and (\ref{35}) that ACI effects cannot occur for two
independent single-photon wavepackets. Therefore, ACI is the signature of
two-photon entanglement. We will discuss ACI in details in the next two
sub-sections.

\subsection{anti-coalescence interference and two-photon transparent state}

The other manner of two-photon interference is just opposite of the
coalescence interference: the coincidence probability at the output of
beamsplitter is greater than that of the absence of interference. In the
extreme case two photons never go together, we call it the perfect ACI.

$Case$ $I:$ According to Eq. (\ref{21}), the necessary and sufficient
condition for the perfect ACI is 
\begin{equation}
C(\omega _1,\omega _2)=-C(\omega _2,\omega _1)  \label{asy}
\end{equation}
in the whole frequency space. We call Eq. (\ref{asy}) the anti-symmetric
two-photon spectrum. Obviously, it satisfies $C(\omega ,\omega )=0$. One can
see immediately from Eq. (\ref{21}) that only the output states for two
photons traveling in different ports remain so that the coincidence
probability is unity. Furthermore, when the anti-symmetric condition (\ref
{asy}) is satisfied, the output state (\ref{21}) is reduced to 
\begin{eqnarray}
|\Phi _{two}\rangle _{out} &=&\sum_{\omega _1<\omega _2}C(\omega _1,\omega
_2)[a_1^{\dagger }(\omega _1)a_2^{\dagger }(\omega _2)-a_1^{\dagger }(\omega
_2)a_2^{\dagger }(\omega _1)]|0\rangle  \label{9p} \\
&=&\sum_{\omega _1<\omega _2}C(\omega _1,\omega _2)a_1^{\dagger }(\omega
_1)a_2^{\dagger }(\omega _2)|0\rangle -\sum_{\omega _2<\omega _1}C(\omega
_2,\omega _1)a_1^{\dagger }(\omega _1)a_2^{\dagger }(\omega _2)]|0\rangle 
\nonumber \\
&=&\sum_{\omega _1,\omega _2}C(\omega _1,\omega _2)a_1^{\dagger }(\omega
_1)a_2^{\dagger }(\omega _2)|0\rangle =|\Phi _{two}\rangle .  \nonumber
\end{eqnarray}
It means that when the perfect ACI occurs, the output state is identical to
the input. In other words, a two-photon wavepacket with the anti-symmetric
spectrum (\ref{asy}) is invariant under the 50/50 beamsplitter transform, or
it is the eigenstate with the unity eigenvalue. Note that the eigenstate is
not unique, and it can be any two-photon wavepacket satisfying condition (%
\ref{asy}). Physically, the two-photon wavepacket is perfectly transparent
passing beamsplitter, so we call it two-photon transparent state. Of course,
the two-photon transparent state must be in entanglement, since two
independent single-photon wavepackets never show ACI effect. A well-known
example of two-photon transparent state is the Bell state $|\Psi ^{-}\rangle 
$ which is defined by (\ref{b1b}) and satisfies the anti-symmetric condition
(\ref{asy}). So the Bell state $|\Psi ^{-}\rangle $ is the eigenstate in a
50/50 beamsplitter transform. This is the reason why the Bell state $|\Psi
^{-}\rangle $ can be measured by a coincidence counting in the teleportation
scheme.\cite{bou}

$Case$ $II:$ Similarly as discussed above, the necessary and sufficient
conditions for the perfect ACI are 
\begin{mathletters}
\label{43}
\begin{eqnarray}
C_{mm}(\omega _1,\omega _2) &\equiv &-C_{mm}(\omega _2,\omega _1),\qquad
m=\alpha ,\beta ,  \label{43a} \\
C_{\alpha \beta }(\omega _1,\omega _2) &\equiv &-C_{\beta \alpha }(\omega
_2,\omega _1).  \label{43b}
\end{eqnarray}
The same as case I, condition (\ref{43a}) gives the invariant state 
\end{mathletters}
\begin{equation}
|\Phi _{mm}\rangle _{out}=|\Phi _{mm}\rangle ,\qquad m=\alpha ,\beta .
\label{44}
\end{equation}
However, under condition (\ref{43b}), one obtains 
\begin{eqnarray}
|\Phi _{\alpha \beta }\rangle _{out} &=&\sum_{\omega _1,\omega _2}C_{\alpha
\beta }(\omega _1,\omega _2)[a_{1\alpha }^{\dagger }(\omega _1)a_{2\beta
}^{\dagger }(\omega _2)-a_{2\alpha }^{\dagger }(\omega _1)a_{1\beta
}^{\dagger }(\omega _2)]|0\rangle  \label{45} \\
&=&\sum_{\omega _1,\omega _2}[C_{\alpha \beta }(\omega _1,\omega
_2)a_{1\alpha }^{\dagger }(\omega _1)a_{2\beta }^{\dagger }(\omega
_2)+C_{\beta \alpha }(\omega _2,\omega _1)a_{2\alpha }^{\dagger }(\omega
_1)a_{1\beta }^{\dagger }(\omega _2)]|0\rangle =|\Phi _{\alpha \beta
}\rangle ,\qquad \alpha \leftrightarrow \beta .  \nonumber
\end{eqnarray}
Again, a two-photon wavepacket with the anti-symmetric spectra (\ref{43}) is
transparent passing the beamsplitter. It is readily to check that the
polarization-entangled anti-symmetric Bell state $|\Psi ^{-}\rangle $ and
state $|\Psi _w(\theta =\pi )\rangle $, defined by Eqs. (\ref{17b}) and (\ref
{18}), respectively, fulfill conditions (\ref{43}) and hence are the
two-photon transparent states.

We have already indicated that, there is no interference among two-photon
pairs, $\alpha \alpha ,\beta \beta $ and $\alpha \beta $, so that conditions
(\ref{43a}) and (\ref{43b}) are for the perfect ACIs of two photons with the
same polarization and the orthogonal polarizations, respectively. For case
II, it is possible that one of the two-photon spectra is symmetric and the
other one is anti-symmetric. For example, we consider two independent
single-photon wavepackets which are identical . Then a phase shift $\theta $
for $\beta -$polarized beam is introduced by inserting a wave-plate in path
1. The combined two-photon state is written as

\begin{eqnarray}
|\Psi _{ss}(\theta )\rangle &=&\sum_{\omega _1}[C_\alpha (\omega
_1)a_{1\alpha }^{\dagger }(\omega _1)+e^{i\theta }C_\beta (\omega
_1)a_{1\beta }^{\dagger }(\omega _1)]\sum_{\omega _2}[C_\alpha (\omega
_2)a_{2\alpha }^{\dagger }(\omega _2)+C_\beta (\omega _2)a_{2\beta
}^{\dagger }(\omega _2)]|0\rangle  \label{46} \\
&=&\sum_{\omega _1,\omega _2}[C_\alpha (\omega _1)C_\alpha (\omega
_2)a_{1\alpha }^{\dagger }(\omega _1)a_{2\alpha }^{\dagger }(\omega
_2)+e^{i\theta }C_\beta (\omega _1)C_\beta (\omega _2)a_{1\beta }^{\dagger
}(\omega _1)a_{2\beta }^{\dagger }(\omega _2)  \nonumber \\
&&+C_\alpha (\omega _1)C_\beta (\omega _2)a_{1\alpha }^{\dagger }(\omega
_1)a_{2\beta }^{\dagger }(\omega _2)+e^{i\theta }C_\beta (\omega _1)C_\alpha
(\omega _2)a_{1\beta }^{\dagger }(\omega _1)a_{2\alpha }^{\dagger }(\omega
_2)]|0\rangle .  \nonumber
\end{eqnarray}
The two-photon spectra of state $|\Psi _{ss}(\theta =\pi )\rangle $ satisfy
the symmetric condition (\ref{33a}) for the photon pairs $\alpha \alpha $
and $\beta \beta $, and the anti-symmetric condition (\ref{43b}) for the
photon pair $\alpha \beta $. In result, the photon pairs with the same
polarization travel together while the photon pairs with the orthogonal
polarizations exit from different ports. Nevertheless, the total coincidence
probability in polarization-insensitive detection must satisfy Eq. (\ref{35}%
).

\subsection{observation of anti-coalescence interference effect}

The detection of entanglement is the one of the important tasks in quantum
information. We have proved in the previous subsections that the ACI effect
is the signature of two-photon entanglement so that it can be an useful and
simple method to demonstrate entanglement.

The photon entanglement state generated in the source may have a symmetric
spectrum showing the CI effect. Due to the fact that the manners of
interference depend on the relative phase of the interference term which
increases or decreases coincidence probability, we introduce an additional
phase in two-photon wavepacket to change the manners of the interference
from CI to ACI.

$Case$ $I:$ We consider a two-photon spectrum in the form of 
\begin{equation}
Q(\omega _1,\omega _2)=g(\omega _1+\omega _2-2\Omega )f(\omega _1-\Omega
)f(\omega _2-\Omega ),  \label{50}
\end{equation}
in which $f(x)$ describes a spectral profile identical for two single-photon
beams and $g(x)$ describes possible entanglement. Obviously, spectrum (\ref
{50}) is symmetric. If $f(x)$ is a Gaussian, we have already calculated the
coincidence probability as shown in Eq. (\ref{32}) which is in fact
irrelevant to photon entanglement. In order to introduce an additional phase
in the spectrum, one can set an unbalanced Mach-Zehnder interferometer in
one path of the beam. This method was proposed in the previous experiments.%
\cite{chiao}\cite{shih5} We explain this method again in the S-picture by
the spectra feature for two-photon state. Let beam 1 be split into two
parts, and one travels a short path $L_s$, and the other a long path $L_l$.
Then these two sub-beams incorporate a beam again which interferes with beam
2 traveling a path $z_2$. The new two-photon spectrum at the input ports of
beamsplitter is obtained as 
\begin{equation}
C(\omega _1,\omega _2)=Q(\omega _1,\omega _2)[e^{i\omega _1L_l/c}+e^{i\omega
_1L_s/c}]e^{i\omega _2z_2/c}=2Q(\omega _1,\omega _2)e^{i(\omega _1z_1+\omega
_2z_2)/c}\cos (\omega _1\Delta L/c),  \label{47}
\end{equation}
where $z_1=(L_l+L_s)/2$ and $\Delta L=(L_l-L_s)/2$. We set $\nu _i=\omega
_i-\Omega ,$ $(i=1,2)$, Eq. (\ref{47}) is written as 
\begin{equation}
C(\nu _1,\nu _2)=2e^{i\Omega (z_1+z_2)/c}Q(\nu _1,\nu _2)e^{i(\nu _1z_1+\nu
_2z_2)/c}\cos (\nu _1\Delta L/c+\theta ),  \label{48}
\end{equation}
where the additional phase $\theta =\Omega \Delta L/c=2\pi \Delta L/\lambda $%
. $\lambda =2\pi c/\Omega $ is the wavelength for each single-photon beam.
In the case of the perfect phase matching in SPDC, $g(x)\rightarrow \delta
(x),$ one obtains 
\begin{mathletters}
\label{49}
\begin{eqnarray}
C(\nu _1,\nu _2) &=&\pm 2e^{i\Omega (z_1+z_2)/c}\delta (\nu _1+\nu _2)f(\nu
_1)f(\nu _2)e^{i(\nu _1z_1+\nu _2z_2)/c}\cos (\nu _1\Delta L/c),\qquad
\theta =n\pi ,  \label{49a} \\
C(\nu _1,\nu _2) &=&\pm 2e^{i\Omega (z_1+z_2)/c}\delta (\nu _1+\nu _2)f(\nu
_1)f(\nu _2)e^{i(\nu _1z_1+\nu _2z_2)/c}\sin (\nu _1\Delta L/c),\qquad
\theta =(n+\frac 12)\pi .  \label{49b}
\end{eqnarray}
At the balanced position $z_1=z_2$, spectrum (\ref{49a}) is symmetric and
spectrum (\ref{49b}) is anti-symmetric so that the phase $\theta $ dominates
the interference manners changed between the perfect CI and ACI.

Now, we consider the two-photon spectrum $Q(\omega _1,\omega _2)$ at the
source defined by Eqs. (\ref{12}) and (\ref{13}). Using Eq. (\ref{23}), we
calculate the coincidence probability for the two-photon spectrum (\ref{48})
(see Appendix B) 
\end{mathletters}
\begin{mathletters}
\label{52}
\begin{eqnarray}
P_c &=&\frac 12\{1-\frac 1B[\cos (2\theta )\cdot e^{-\frac 12(\frac{\beta ^2%
}{2+\beta ^2}\Delta L^2+\Delta z^2)(\frac \sigma c)^2}+\frac 12e^{-\frac 12%
(\Delta L+\Delta z)^2(\frac \sigma c)^2}+\frac 12e^{-\frac 12(\Delta
L-\Delta z)^2(\frac \sigma c)^2}]\},  \label{52a} \\
B &=&1+\cos (2\theta )\exp [-\frac{1+\beta ^2}{2+\beta ^2}\Delta L^2(\frac 
\sigma c)^2],  \label{52b}
\end{eqnarray}
where $\beta \equiv \sigma _p/\sigma $ and $\Delta z=z_2-z_1$. The three
terms in the square brackets of Eq. (\ref{52a}) contribute to the
interference occurring mainly at the three positions of beamsplitter: the
first term for $\Delta z=0$ and the last two terms for $\Delta z=\pm \Delta
L $. Similar to Eq. (\ref{32}), the coherent length of the single-photon
beam $c/\sigma $ defines the width of the interference dip (or peak) so that
only when $\Delta L$ is larger than $c/\sigma $ the dips can be apart in
space. In the first term, the phase $2\theta $ may affect the interference
manners, CI, ACI or the absence of interference, whereas in the last two
terms it shows only the CI effect. However, to show a significant
interference effect at the balanced position, it should satisfy the
condition 
\end{mathletters}
\begin{equation}
\Delta L<\sqrt{\frac{2+\beta ^2}{\beta ^2}}\frac c\sigma =\sqrt{2+\beta ^2}%
\frac c{\sigma _p},  \label{53m}
\end{equation}
where $c/\sigma _p$ is the coherent length for the pump beam. Since $\sigma
_p$ is related to two-photon entanglement, $c/\sigma _p$ is also called the
two-photon coherent length. For $\sigma _p<<\sigma $ ($\beta <<1$), that is
the two-photon coherent length is much larger than the single-photon
coherent length, it is possible that the optical path difference of two
beams $\Delta L$ exceeds the single-photon coherent length $c/\sigma $, but
condition (\ref{53m}) is satisfied. This fact has been demonstrated
experimentally in Ref. \cite{shih5}.

For the perfect phase matching in SPDC, $g(x)=\delta (x)$ is set in Eq. (\ref
{12}), the coincidence probability can be calculated by Eq. (\ref{23}) 
\begin{equation}
P_c=\frac 12\{1-\frac 1{1+\cos 2\theta \cdot e^{-\frac 12\Delta L^2(\frac 
\sigma c)^2}}[\cos 2\theta \cdot e^{-\frac 12\Delta z^2(\frac \sigma c)^2}+%
\frac 12e^{-\frac 12(\Delta L+\Delta z)^2(\frac \sigma c)^2}+\frac 12e^{-%
\frac 12(\Delta L-\Delta z)^2(\frac \sigma c)^2}]\}.  \label{53}
\end{equation}
If $\Delta L>>c/\sigma $, the above equation is approximately written as 
\begin{equation}
P_c\approx \frac 12\{1-\cos (2\theta )e^{-\frac 12\Delta z^2(\frac \sigma c%
)^2}-\frac 12e^{-\frac 12(\Delta L+\Delta z)^2(\frac \sigma c)^2}-\frac 12%
e^{-\frac 12(\Delta L-\Delta z)^2(\frac \sigma c)^2}]\}.  \label{54}
\end{equation}
This result was obtained in the previous study\cite{shih5}. (Note that Eq. (%
\ref{54}) is approximately valid since it gives $P_c<0$ at $\Delta z=0$ for $%
\theta =n\pi $.) At the balanced position, Equation (\ref{53}) is simplified
as 
\begin{equation}
P_c=\frac{(1-\cos 2\theta )[1-e^{-\frac 12\Delta L^2(\frac \sigma c)^2}]}{%
2[1+\cos 2\theta \cdot e^{-\frac 12\Delta L^2(\frac \sigma c)^2}]}.
\label{51}
\end{equation}
It shows that, for an ideal two-photon entanglement in frequency, the
perfect CI and ACI occur by setting the phase $\theta =n\pi $ and $\theta
=(n+1/2)\pi $, respectively. This is consistent with the symmetry of
two-photon spectrum indicated by Eq. (\ref{49}).

To show the feature of the entanglement, we also apply this method to two
independent single-photon wavepackets for comparison. In this case, we set $%
g(x)$ constant in Eq. (\ref{12}), and the two single-photon spectra are
separable as 
\begin{eqnarray}
C_1(\nu ) &=&A_1e^{-\nu ^2/(2\sigma ^2)}e^{i\nu z_1/c}\cos (\nu \Delta
L/c+\theta ),  \label{56} \\
C_2(\nu ) &=&A_2e^{-\nu ^2/(2\sigma ^2)}e^{i\nu z_2/c},  \nonumber
\end{eqnarray}
in which the phase factor independent of frequency is neglected. By using
Eq. (\ref{30p}), we calculate the coincidence probability for the above two
independent single-photon spectra (see Appendix C) 
\begin{eqnarray}
P_c &=&\frac 12\{1-\frac 1{2(1+\cos 2\theta \cdot e^{-\Delta L^2(\frac \sigma
c)^2})}|e^{-i\theta }e^{-\frac 14(\Delta L+\Delta z)^2(\frac \sigma c%
)^2}+e^{i\theta }e^{-\frac 14(\Delta L-\Delta z)^2(\frac \sigma c)^2}|^2\}
\label{55} \\
&=&\frac 12\{1-\frac 1{1+\cos 2\theta \cdot e^{-\Delta L^2(\frac \sigma c)^2}%
}[\cos 2\theta \cdot e^{-\frac 12(\Delta L^2+\Delta z^2)(\frac \sigma c)^2}+%
\frac 12e^{-\frac 12(\Delta L+\Delta z)^2(\frac \sigma c)^2}+\frac 12e^{-%
\frac 12(\Delta L-\Delta z)^2(\frac \sigma c)^2}]\}.  \nonumber
\end{eqnarray}
The first line of Eq. (\ref{55}) shows clearly $P_c<1/2$. We note that the
same results as Eqs. (\ref{53}) and (\ref{55}) can also be obtained from Eq.
(\ref{52}) by setting $\beta \rightarrow 0$ and $\beta \rightarrow \infty $,
respectively.

In Figs. 3-5, we plot the coincidence probabilities for the three examples
of two-photon spectra: the maximum two-photon entanglement described by $%
g(x)=\delta (x)$, the arbitrary entanglement described by Eq. (\ref{13})
with $\sigma _p=\sigma $, and the two independent single-photon wavepackets (%
\ref{56}), which are indicated by solid, dashed and dotted lines,
respectively. Figures 3a and 3b show the coincidence probability versus the
phase $2\theta $ at the balanced position $z_1=z_2$ for $\Delta L(\sigma
/c)= $1 and 3, respectively. It shows that the phase $\theta $ dominates the
manners of interference. For the maximum entanglement (solid line), the
perfect CI and ACI occur at the phase $2\theta =2n\pi $ and $(2n+1)\pi $,
respectively, and it is independent of the normalized optical path
difference $\Delta L(\sigma /c).$ For an arbitrary entanglement, however,
both CI and ACI can occur, but not perfect (dashed line). As for two
independent single-photon wavepackets in the coherent range for
single-photon ($\Delta L(\sigma /c)=1$), the CI occurs, but there is no ACI
effect (dotted line).

In experiments, it would be difficult to measure the net coincidence
probability due to a lower quantum efficiency. The curves in Fig. 3 are
unable to witness ACI\ effect if there is not a\ reference for coincidence
probability. Alternatively, one may scan the position of beamsplitter to
show the two-photon interference. In Fig. 4, we plot the coincidence
probability versus the normalized position of the beamsplitter $\Delta
z(\sigma /c)$ for $\Delta L(\sigma /c)=1$. The reference of coincidence
probability has been shown at the large $\Delta z(\sigma /c)$. In Fig. 4a,
by choosing the phase $2\theta =(2n+1)\pi $, the ACI effect has been shown
at the balanced position, and it witnesses the two-photon entanglement of
the input state. In Fig. 4b, $2\theta =2n\pi $, the CI occurs, and there is
no significant difference for the three cases. In Fig. 4c (also in Fig. 5c)
for $2\theta =(2n+1/2)\pi $, however, the three curves coincide exactly,
showing the CI. As a matter of fact, for $2\theta =(2n+1/2)\pi $, Eqs.\ (\ref
{52}), (\ref{53}) and (\ref{55}) become identical 
\begin{equation}
P_c=\frac 12\{1-\frac 12[e^{-\frac 12(\Delta L+\Delta z)^2(\frac \sigma c%
)^2}+e^{-\frac 12(\Delta L-\Delta z)^2(\frac \sigma c)^2}]\}.  \label{old}
\end{equation}
In this case, the interference is independent of photon entanglement
evaluated by the bandwidth $\sigma _p$ of the pump beam. In Fig. 5, we set a
larger $\Delta L(\sigma /c)=3,$ for which the traveling path difference $%
\Delta L$ of two photons is larger than the coherent length $c/\sigma $ of
the single-photon beam. The two side-dips emerge approximately at the
position $\Delta z=\Delta L$. Different from Fig. 4, for two independent
single-photon wavepackets (shown by the dotted lines), the interferences
disappear at the balanced position for any value of phase $\theta $. But the
CI and ACI still occur for the entangled two-photon wavepacket by choosing
proper phases.

The interference effect shown in Fig. 5 has been reported in Ref. \cite
{shih5}, in which the authors demonstrate that ''the single-photon
wavepacket concept is not always appropriate for two-photon interference
measurements''. Due to the fact that in the experiment two photons to be
interfered are in entanglement the Feynman-type diagrams of biphoton
amplitudes were applied in their theoretical analysis. In the present theory
we show an uniform description for a general wavepacket containing two
photons, whether in entanglement or not. It has shown that the entangled
two-photon wavepacket may behave in interference manners similar to or
different from the un-entangled one. But only ACI effect is the signature of
two-photon entanglement. The various effects can be understood by the
two-photon spectra which are typically described by Eqs. (\ref{48}), (\ref
{50}), (\ref{12}) and (\ref{13}). We plot the contours for the envelope of
the spectra by omitting the oscillatory phase factor (i.e. setting $%
z_1=z_2=0 $ in Eq. (\ref{48})) for simplicity. Each point in the spectral
plane corresponds to a two-photon state with the amplitude $C(\nu _1,\nu _2)$%
. In the contour, pairs of points symmetric with respect to the diagonal $%
\nu _1=\nu _2$ contribute to interference so that the topological
characteristic of contour may illustrate the interference manners. Figures 6
and 7 show the contours of the spectra for the un-entangled (by setting $%
g(x)=1$) and entangled (by setting $\sigma _p=(1/3)\sigma $) wavepackets,
respectively, in which the bright and the dark with respect to the
background indicate respectively the positive and negative values of
amplitudes. In Figs. 6 and 7, the parameters are chosen as (a) $\Delta
L(\sigma /c)=1$ and $\theta =n\pi ;$ (b) $\Delta L(\sigma /c)=1$ and $\theta
=(n+1/2)\pi ;$ (c) $\Delta L(\sigma /c)=3$ and $\theta =n\pi $; (d) $\Delta
L(\sigma /c)=3$ and $\theta =(n+1/2)\pi $. For un-entangled two-photon
wavepacket, in Fig. 6, the contours of spectra are symmetric with respect to
the Cartesian axis, but not to the diagonal. As for entangled two-photon
wavepacket, the contours in Fig. 7 show approximate symmetry with respect to
the diagonal: the symmetric for Figs. 7a and 7c and the anti-symmetric for
Figs. 7b and 7d.

$Case$ $II:$ We discuss two examples: one is the polarization entangled
two-photon wavepacket described by Eq. (\ref{18}), and the other one
consists of two independent single-photon wavepackets being in two
orthogonally polarized modes described by Eq. (\ref{46}). The manners of
two-photon interference depends mainly on the phase factor $e^{i\theta }$,
which can be set as desire by inserting a wave-plate in path 1 for $\beta $%
-polarization.

To show the reference of interference, let beam $j$ travel a path $z_j$
before entering beamsplitter. At the input ports of beamsplitter, the
two-photon entangled spectra for Eq. (\ref{18}) are given by 
\begin{mathletters}
\label{58}
\begin{eqnarray}
C_{\alpha \beta }(\omega _1,\omega _2) &=&g(\omega _1+\omega _2-\Omega
_p)e^{-(\omega _1-\Omega _\alpha )^2/(2\sigma _\alpha ^2)-(\omega _2-\Omega
_\beta )^2/(2\sigma _\beta ^2)}e^{i(\omega _1z_1+\omega _2z_2)/c},
\label{58a} \\
C_{\beta \alpha }(\omega _1,\omega _2) &=&e^{i\theta }g(\omega _1+\omega
_2-\Omega _p)e^{-(\omega _1-\Omega _\beta )^2/(2\sigma _\beta ^2)-(\omega
_2-\Omega _\alpha )^2/(2\sigma _\alpha ^2)}e^{i(\omega _1z_1+\omega
_2z_2)/c}.  \label{58b}
\end{eqnarray}
First, we assume $\sigma _\alpha =\sigma _\beta =\sigma $, and $g(x)$ can be
any form. According to Eq. (\ref{29t}), one obtains the coincidence
probability (see Appendix D) 
\end{mathletters}
\begin{equation}
P_c=2P_c^{\alpha \beta }=\frac 12\{1-\cos [(\Omega _\beta -\Omega _\alpha )%
\frac{\Delta z}c-\theta ]\cdot e^{-\frac 12\Delta z^2(\sigma /c)^2}\}.
\label{57}
\end{equation}
Different from Eq. (\ref{32}), which describes interference of two-photon
wavepacket in the same polarization, the frequency difference of two
orthogonally polarized photons introduces a spatial quantum beating.\cite{ou}%
\cite{ra} For the two-photon wavepacket with only frequency entanglement
generated in SPDC of type I, the difference of the central frequencies of
two converted single-photon beams diminishes the symmetry of the two-photon
spectrum (see Eq. (\ref{31n})). However, for the polarization entangled
two-photon wavepacket generated in SPDC of type II, at the balanced position 
$z_1=z_2$ and $\theta =0,$ the symmetry between $C_{\alpha \beta }(\omega
_1,\omega _2)$ and $C_{\beta \alpha }(\omega _1,\omega _2)$ in Eq. (\ref{58}%
) is maintained for two colored converted beams. At unbalanced positions,
the frequency difference causes a phase shift just like phase $\theta $ and
results in a spatial modulation.

Then we consider $\sigma _\alpha \neq \sigma _\beta $. In order to calculate
coincidence probability, function $g(x)$ has to be defined, for example, by
Eq. (\ref{13}). The coincidence probability is the same as Eq. (\ref{57}),
but $\sigma $ is defined by an effective bandwidth (see Appendix D) 
\begin{equation}
\sigma =\sqrt{\frac{\sigma _p^2\sigma _\alpha ^2+\sigma _p^2\sigma _\beta
^2+4\sigma _\alpha ^2\sigma _\beta ^2}{2(\sigma _p^2+\sigma _\alpha
^2+\sigma _\beta ^2)}}.  \label{61}
\end{equation}
This means that in the polarization-entangled state the frequency
correlation described by function $g(x)$ is insignificant for affecting the
manners of the interference.

For two independent single-photon wavepackets described by Eq. (\ref{46}),
let beam $j$ traveling a path $z_j$, the spectra at the beamsplitter are
obtained as 
\begin{mathletters}
\label{59}
\begin{eqnarray}
C_{1\alpha }(\omega ) &=&A_\alpha e^{-(\omega -\Omega _\alpha )^2/(2\sigma
_\alpha ^2)}e^{i\omega z_1/c},\qquad C_{1\beta }(\omega )=A_\beta e^{i\theta
}e^{-(\omega -\Omega _\beta )^2/(2\sigma _\beta ^2)}e^{i\omega z_1/c},
\label{59a} \\
C_{2\alpha }(\omega ) &=&A_\alpha e^{-(\omega -\Omega _\alpha )^2/(2\sigma
_\alpha ^2)}e^{i\omega z_2/c},\qquad C_{2\beta }(\omega )=A_\beta
e^{-(\omega -\Omega _\beta )^2/(2\sigma _\beta ^2)}e^{i\omega z_2/c},
\label{59b}
\end{eqnarray}
where the normalization is given by 
\end{mathletters}
\begin{equation}
n_\alpha =\sqrt{\pi }\sigma _\alpha A_\alpha ^2,\quad n_\beta =\sqrt{\pi }%
\sigma _\beta A_\beta ^2,\quad n_\alpha +n_\beta =1.  \label{62}
\end{equation}
According to Eqs. (\ref{42}) and (\ref{41}), the coincidence probabilities
for the photons with the same polarization and the cross polarizations are
respectively calculated as 
\begin{eqnarray}
P_c^{mm} &=&\frac 12[n_m^2-|A_m^2\int_{-\infty }^\infty d\omega \cdot
e^{-(\omega -\Omega _p)^2/\sigma _p^2}e^{-i\omega \Delta z/c}|^2]=\frac 12%
[n_m^2-|A_m^2\int_{-\infty }^\infty d\nu \cdot e^{-\nu ^2/\sigma
_p^2}e^{-i\nu \Delta z/c}|^2]  \label{63} \\
&=&\frac 12[n_m^2-A_m^4\pi \sigma _p^2e^{-\frac 12\Delta z^2(\sigma
_p/c)^2}]=\frac 12n_m^2[1-e^{-\frac 12\Delta z^2(\sigma _p/c)^2}],\qquad
m=\alpha ,\beta  \nonumber
\end{eqnarray}
and 
\begin{eqnarray}
P_c^{\alpha \beta } &=&P_c^{\beta \alpha }=\frac 12n_\alpha n_\beta -\frac 14%
[A_\alpha ^2\int_{-\infty }^\infty d\omega \cdot e^{-(\omega -\Omega _\alpha
)^2/\sigma _\alpha ^2}e^{-i\omega \Delta z/c}\times A_\beta ^2e^{-i\theta
}\int_{-\infty }^\infty d\omega \cdot e^{-(\omega -\Omega _\beta )^2/\sigma
_\beta ^2}e^{i\omega \Delta z/c}+\text{c.c.}]  \label{64} \\
&=&\frac 12n_\alpha n_\beta -\frac 14[A_\alpha ^2A_\beta ^2\pi \sigma
_\alpha \sigma _\beta e^{i[(\Omega _\beta -\Omega _\alpha )\Delta z/c-\theta
]}e^{-\frac 14\Delta z^2(\sigma _\alpha ^2+\sigma _\beta ^2)/c^2}+\text{c.c.}%
]  \nonumber \\
&=&\frac 12n_\alpha n_\beta \{1-\cos [(\Omega _\beta -\Omega _\alpha )\Delta
z/c-\theta ]\cdot e^{-\frac 14\Delta z^2(\sigma _\alpha ^2+\sigma _\beta
^2)/c^2}\}.  \nonumber
\end{eqnarray}
The total coincidence probability is given by 
\begin{eqnarray}
P_c &=&P_c^{\alpha \alpha }+P_c^{\beta \beta }+2P_c^{\alpha \beta }
\label{65} \\
&=&\frac 12\{1-n_\alpha ^2e^{-\frac 12\Delta z^2(\sigma _\alpha
/c)^2}-n_\beta ^2e^{-\frac 12\Delta z^2(\sigma _\beta /c)^2}-2n_\alpha
n_\beta \cos [(\Omega _\beta -\Omega _\alpha )\Delta z/c-\theta ]\cdot e^{-%
\frac 14\Delta z^2(\sigma _\alpha ^2+\sigma _\beta ^2)/c^2}\}  \nonumber \\
&=&\frac 12\{1-|n_\alpha e^{-\frac 14\Delta z^2(\sigma _\alpha
/c)^2}e^{-i\Omega _\alpha \Delta z/c+i\frac 12\theta }+n_\beta e^{-\frac 14%
\Delta z^2(\sigma _\beta /c)^2}e^{-i\Omega _\beta \Delta z/c-i\frac 12\theta
}|^2\}.  \nonumber
\end{eqnarray}
In the case of $\sigma _\alpha =\sigma _\beta \equiv \sigma $ and $n_\alpha
=n_\beta =1/2$, Eq. (\ref{65}) is simplified as 
\begin{equation}
P_c=\frac 12\{1-\frac 12(1+\cos [(\Omega _\beta -\Omega _\alpha )\frac{%
\Delta z}c-\theta ]\cdot e^{-\frac 12\Delta z^2(\sigma /c)^2})\}.
\label{65p}
\end{equation}

In Figs. 8-12, we plot the coincidence probabilities for the above two
examples of case II, the polarization entangled and un-entangled two-photon
wavepackets, described by Eqs. (\ref{58}) and (\ref{59}), respectively.
First, we consider the coincidence probability at the balanced position $%
z_1=z_2$ of the beamsplitter where the interference effect is significant.
For the entangled wavepacket, Eq. (\ref{57}) becomes 
\begin{equation}
P_c=(1/2)(1-\cos \theta ).  \label{66}
\end{equation}
It verifies the previous discussion: for the spectrum with the symmetry the
perfect CI and ACI effects occur by setting phase $\theta =0$ and $\pi $,
respectively. For the un-entangled wavepacket, according to Eq. (\ref{63}),
the pairs with the same polarization show the perfect CI, $P_c^{\alpha
\alpha }=P_c^{\beta \beta }=0$ at the balanced position. In result, the
total coincidence probability is obtained as 
\begin{equation}
P_c=2P_c^{\alpha \beta }=n_\alpha n_\beta (1-\cos \theta ).  \label{67}
\end{equation}
Similarly, it can perform the perfect CI at $\theta =0$. Because of the
normalization (\ref{62}), it has $n_\alpha n_\beta \leq 1/4$. Hence the
maximum coincidence probability in Eq. (\ref{67}) is $P_c=1/2$ for $\theta
=\pi $. Figure 8 shows the total coincidence probabilities versus phase $%
\theta $ for these two examples. Note that, for the two modes case, $P_c=1/2$
does not always mean ''the absence of interference'', and it will be
illustrated in Fig. 11.

For the sake of showing the reference, the coincidence probabilities versus
the normalized position of beamsplitter are plotted in Figs. 9-12, in which
Figs. 9 and 10 are for the polarization entangled wavepacket, and Figs. 11
and 12 for the two independent single-photon wavepackets. In Fig. 9, by
setting $\Omega _\alpha =\Omega _\beta $, it shows the different profiles of
the interference, depending on phase $\theta $. The observable ACI effect
for the phase $\theta =\pi $ shows the evidence for two-photon entanglement.
It is interesting that, when $\theta =\pi /2$, there is no interference
completely. This is because of out of phase for two amplitudes of the
two-photon states interfered. When $\Omega _\alpha \neq \Omega _\beta $, the
coincidence probabilities display the interference fringe shown in Fig. 10.
The phase causes the shift of the fringe.

For two independent single-photon wavepackets being in two polarization
modes, if the detection system can recognize the polarization, one can
observe polarization-sensitive two-photon interferences. In Fig. 11, in
which $n_\alpha =n_\beta =1/2,$ $\Omega _\alpha =\Omega _\beta $ and $\sigma
_\alpha =\sigma _\beta \equiv \sigma $ are set in Eqs. (\ref{59}) and (\ref
{62}), it shows that the photon pair with the same polarization $(\alpha
\alpha )$ or $(\beta \beta )$ perform the same CI which is independent of
phase $\theta $, while the photon pair with the orthogonal polarizations may
show different manners of interferences, depending on phase $\theta $. For
example, when $\theta =\pi $, the interference pattern for $P_c^{\alpha
\alpha }$ or $P_c^{\beta \beta }$ shows the same dip, whereas for $%
P_c^{\alpha \beta }+P_c^{\beta \alpha }$ it shows a peak. In result, the
total coincidence probability satisfies $P_c\equiv 1/2$. Physically, it does
not mean ''the absence of interference'', because the two opposite manners
of interferences do occur. The photons with the same polarization travel
together while the photons with the different polarizations travel apart.
The peak observed in the polarization-sensitive detection does not mean that
the input two-photon wavepacket is in entanglement. In order to demonstrate
the entanglement, one must measure the coincidence probability insensitive
to polarization. When the difference of the central frequencies for two
polarized modes is introduced, $\Omega _\alpha \neq \Omega _\beta $, the
interference fringes appear as shown in Fig. 12. In comparison of Fig. 12
with Fig. 10, it shows clearly the difference of the reference level in the
interference fringes, which witnesses the two-photon entanglement.

\section{Conclusion}

In conclusion, we study two-photon interference for a general two-photon
wavepacket with a finite spectral range in the representation of the quantum
state. It is clearly shown that two-photon interference originates from the
indistinguishability of two-photon states, whether for input an entangled
two-photon wavepacket or two independent single-photon wavepackets. Various
behaviors of two-photon interferences can be understood by the topological
symmetry of two-photon spectral amplitude. We distinguish the CI and ACI
effects according to the coincidence probability less and more than that for
the absence of interference. We prove that un-entangled two-photon
wavepackets never show ACI effect, so it makes possibility to witness the
photon entanglement by the ACI effect. However, the necessary and sufficient
conditions for the perfect CI and ACI are deduced. For a two-photon
wavepacket with anti-symmetric spectrum, the perfect ACI occurs and the
wavepacket passes a 50/50 beamsplitter transparently.

In this paper, we consider the two-photon wavepacket propagating in one
dimension. Without difficulty, the present method can be extended to discuss
the beam with transverse distribution. The recent work\cite{wa} has shown
that the spatial symmetry of wavefunction can also affect the interference
manners.

\section{Acknowledgment}

The authors thank Sh. Y. Zhu and G. X. Li for discussions. This research was
supported by the National Program of Fundamental Research No. 2001CB309310
and the National Natural Science Foundation of China, Project Nos. 10074008
and 60278021.

\section{Appendices}

\subsection{Appendix A}

Consider the two-photon spectrum (\ref{31n}), the normalization is given by 
\begin{eqnarray}
1 &=&\int_{-\infty }^\infty d\omega _1\int_{-\infty }^\infty d\omega
_2|g(\omega _1+\omega _2-\Omega _p)|^2e^{-[(\omega _1-\Omega _1)^2+(\omega
_2-\Omega _2)^2]/\sigma ^2}  \eqnum{A1}  \label{A1} \\
&=&\int_{-\infty }^\infty d\nu _1\int_{-\infty }^\infty d\nu _2|g(\nu _1+\nu
_2)|^2e^{-[\nu _1^2+\nu _2^2]/\sigma ^2},  \nonumber
\end{eqnarray}
where $\nu _i=\omega _i-\Omega _i$ $(i=1,2)$. Let 
\begin{equation}
\nu _p=\nu _1+\nu _2,\qquad \nu _m=\nu _2-\nu _1,  \eqnum{A2}  \label{A2}
\end{equation}
Equation (\ref{A1}) is obtained as 
\begin{equation}
1=\frac 12\int_{-\infty }^\infty d\nu _p\int_{-\infty }^\infty d\nu _m|g(\nu
_p)|^2e^{-[\nu _p^2+\nu _m^2]/(2\sigma ^2)}=\sigma \sqrt{\frac \pi 2}%
\int_{-\infty }^\infty d\nu _p|g(\nu _p)|^2e^{-\nu _p^2/(2\sigma ^2)}. 
\eqnum{A3}  \label{A3}
\end{equation}
Then, we calculate the integration by taking into account Eq. (\ref{31}) 
\begin{eqnarray}
&&\frac 12\int_{-\infty }^\infty d\omega _1\int_{-\infty }^\infty d\omega
_2[C(\omega _1,\omega _2)C^{*}(\omega _2,\omega _1)+\text{c.c.}]  \eqnum{A4}
\label{A4} \\
&=&\int_{-\infty }^\infty d\omega _1\int_{-\infty }^\infty d\omega
_2|g(\omega _1+\omega _2-\Omega _p)|^2e^{-[(\omega _1-\Omega _1)^2+(\omega
_2-\Omega _2)^2+(\omega _1-\Omega _2)^2+(\omega _2-\Omega _1)^2]/(2\sigma
^2)}\cos [(\omega _2-\omega _1)\Delta z/c]  \nonumber \\
&=&\int_{-\infty }^\infty d\nu _1\int_{-\infty }^\infty d\nu _2|g(\nu _1+\nu
_2)|^2e^{-[(\nu _1{}^2+(\nu _2+\Delta \Omega )^2+\nu _2{}^2+(\nu _1-\Delta
\Omega )^2]/(2\sigma ^2)}\cos [(\nu _2-\nu _1+\Delta \Omega )\Delta z/c] 
\nonumber \\
&=&\frac 12\int_{-\infty }^\infty d\nu _p\int_{-\infty }^\infty d\nu
_m|g(\nu _p)|^2e^{-[(\nu _p^2+\nu _m^2)+2\nu _m\Delta \Omega ]/(2\sigma
^2)}e^{-(\Delta \Omega /\sigma )^2}\cos [(\nu _m+\Delta \Omega )\Delta z/c] 
\nonumber \\
&=&\frac 12e^{-\frac 12(\Delta \Omega /\sigma )^2}\int_{-\infty }^\infty
d\nu _p|g(\nu _p)|^2e^{-\nu _p^2/(2\sigma ^2)}\int_{-\infty }^\infty d\nu
_me^{-(\nu _m+\Delta \Omega )^2/(2\sigma ^2)}\cos [(\nu _m+\Delta \Omega
)\Delta z/c]  \nonumber \\
&=&e^{-\frac 12(\sigma \Delta z/c)^2}e^{-\frac 12(\Delta \Omega /\sigma )^2}.
\nonumber
\end{eqnarray}
According to Eq. (\ref{23}), one obtains Eq. (\ref{32}). Similarly, Eq. (\ref
{32pc}) can be calculated.

\subsection{Appendix B}

Consider a two-photon spectrum described by Eqs. (\ref{48}), in which $Q(\nu
_1,\nu _2)=(A/2)e^{-(\nu _1+\nu _2)^2/2\sigma _p^2}e^{-(\nu _1^2+\nu
_2^2)/2\sigma ^2}$. The normalization is given by 
\begin{eqnarray}
1 &=&\int_{-\infty }^\infty d\nu _1\int_{-\infty }^\infty d\nu _2|C(\nu
_1,\nu _2)|^2=A^2\int_{-\infty }^\infty d\nu _1\int_{-\infty }^\infty d\nu
_2\cdot e^{-(\nu _1+\nu _2)^2/\sigma _p^2}e^{-(\nu _1^2+\nu _2^2)/\sigma
^2}\cos ^2(\nu _1\Delta L/c+\theta )  \eqnum{B1}  \label{B1} \\
&=&A^2\int_{-\infty }^\infty d\nu _1\int_{-\infty }^\infty d\nu _2\cdot
e^{-(\nu _1+\nu _2)^2/\sigma _p^2}e^{-(\nu _1^2+\nu _2^2)/\sigma ^2}\frac 12%
[1+\cos (2\nu _1\Delta L/c)\cos 2\theta ]  \nonumber \\
&=&\frac 14A^2\int_{-\infty }^\infty d\nu _p\int_{-\infty }^\infty d\nu
_m\cdot e^{-\nu _p{}^2/\sigma _p^2}e^{-(\nu _p^2+\nu _m^2)/(2\sigma
^2)}\{1+\cos 2\theta \cos [(\nu _p+\nu _m)\Delta L/c]\}  \nonumber \\
&=&\frac 14A^2\int_{-\infty }^\infty d\nu _p\int_{-\infty }^\infty d\nu
_m\cdot e^{-\nu _p{}^2[1/\sigma _p^2+1/(2\sigma ^2)]}e^{-\nu _m^2/(2\sigma
^2)}\{1+\cos 2\theta \cos (\nu _p\Delta L/c)\cos (\nu _m\Delta L/c)\} 
\nonumber \\
&=&\frac 12A^2\frac{\pi \sigma _p\sigma ^2}{\sqrt{2\sigma ^2+\sigma _p^2}}B,
\nonumber
\end{eqnarray}
where Eq. (\ref{A2}) has been applied, and $B$ is defined by Eq. (\ref{52b}%
). Then, we calculate the integration 
\begin{eqnarray}
&&\int_{-\infty }^\infty d\nu _1\int_{-\infty }^\infty d\nu _2[C(\nu _1,\nu
_2)C^{*}(\nu _2,\nu _1)+\text{c.c.}]  \eqnum{B2}  \label{B2} \\
&=&2A^2\int_{-\infty }^\infty d\nu _1\int_{-\infty }^\infty d\nu _2\cdot
e^{-(\nu _1+\nu _2)^2/\sigma _p^2}e^{-(\nu _1^2+\nu _2^2)/\sigma ^2}\cos
(\nu _1\Delta L/c+\theta )\cos (\nu _2\Delta L/c+\theta )\cos [(\nu _2-\nu
_1)\Delta z/c]  \nonumber \\
&=&A^2\int_{-\infty }^\infty d\nu _1\int_{-\infty }^\infty d\nu _2\cdot
e^{-(\nu _1+\nu _2)^2/\sigma _p^2}e^{-(\nu _1^2+\nu _2^2)/\sigma ^2}\cos
[(\nu _2-\nu _1)\Delta z/c]\{\cos [(\nu _1+\nu _2)\Delta L/c+2\theta ]+\cos
[(\nu _1-\nu _2)\Delta L/c]\}  \nonumber \\
&=&\frac 12A^2\int_{-\infty }^\infty d\nu _p\int_{-\infty }^\infty d\nu
_m\cdot e^{-\nu _p^2/\sigma _p^2}e^{-(\nu _p^2+\nu _m^2)/(2\sigma ^2)}\cos
(\nu _m\Delta z/c)[\cos (\nu _p\Delta L/c+2\theta )+\cos (\nu _m\Delta L/c)]
\nonumber \\
&=&A^2\frac{\pi \sigma _p\sigma ^2}{\sqrt{2\sigma ^2+\sigma _p^2}}\{\cos
2\theta \cdot e^{-\frac 12(\frac{\sigma _p^2}{\sigma _p^2+2\sigma ^2}\Delta
L^2+\Delta z^2)(\frac \sigma c)^2}+\frac 12e^{-\frac 12(\Delta L+\Delta z)^2(%
\frac \sigma c)^2}+\frac 12e^{-\frac 12(\Delta L-\Delta z)^2(\frac \sigma c%
)^2}\}.  \nonumber
\end{eqnarray}
By taking into account the normalization (\ref{B1}), we obtain the
coincidence probability (\ref{52}).

\subsection{Appendix C}

For two independent single-photon spectra defined by Eq. (\ref{56}), the
normalization is given by 
\begin{equation}
1=A_1^2A_2^2\int_{-\infty }^\infty e^{-\nu ^2/\sigma ^2}d\nu \int_{-\infty
}^\infty e^{-\nu ^2/\sigma ^2}\cos ^2(\nu \Delta L/c+\theta )d\nu =A_1^2A_2^2%
\frac 12\pi \sigma ^2(1+\cos 2\theta \cdot e^{-\Delta L^2(\sigma /c)^2}). 
\eqnum{C1}  \label{C1}
\end{equation}
Then, we calculate the integration 
\begin{eqnarray}
\int_{-\infty }^\infty C_1(\nu )C_2^{*}(\nu )d\nu &=&A_1A_2\int_{-\infty
}^\infty e^{-\nu ^2/\sigma ^2}e^{i\nu (z_1-z_2)/c}\cos (\nu \Delta
L/c+\theta )d\nu  \eqnum{C2}  \label{C2} \\
&=&A_1A_2\int_{-\infty }^\infty e^{-\nu ^2/\sigma ^2}\{e^{-i\theta }\cos
[\nu (\Delta L+\Delta z)/c]+e^{i\theta }\cos [\nu (\Delta L-\Delta
z)/c]\}d\nu  \nonumber \\
&=&\frac 12A_1A_2\sqrt{\pi }\sigma \{e^{-i\theta }e^{-\frac 14(\Delta
L+\Delta z)^2(\frac \sigma c)^2}+e^{i\theta }e^{-\frac 14(\Delta L-\Delta
z)^2(\frac \sigma c)^2}\}.  \nonumber
\end{eqnarray}
By using Eq. (\ref{30p}), the coincidence probability is written as 
\begin{equation}
P_c=\frac 12[1-\frac 14A_1^2A_2^2\pi \sigma ^2|e^{-i\theta }e^{-\frac 14%
(\Delta L+\Delta z)^2(\frac \sigma c)^2}+e^{i\theta }e^{-\frac 14(\Delta
L-\Delta z)^2(\frac \sigma c)^2}|^2].  \eqnum{C3}  \label{C3}
\end{equation}
By taking into account the normalization (\ref{C1}), Eq. (\ref{55}) has been
obtained.

\subsection{Appendix D}

First, we consider the case of $\sigma _\alpha =\sigma _\beta =\sigma $. For
the polarization entangled two-photon spectrum defined by Eq. (\ref{58}),
the normalization is given by 
\begin{eqnarray}
1 &=&\int_{-\infty }^\infty d\omega _1\int_{-\infty }^\infty d\omega
_2|C_{\alpha \beta }(\omega _1,\omega _2)|^2+\int_{-\infty }^\infty d\omega
_1\int_{-\infty }^\infty d\omega _2|C_{\beta \alpha }(\omega _1,\omega _2)|^2
\eqnum{D1}  \label{D1} \\
&=&2\int_{-\infty }^\infty d\nu _1\int_{-\infty }^\infty d\nu _2\cdot |g(\nu
_1+\nu _2)|^2e^{-(\nu _1^2+\nu _2^2)/\sigma ^2}=\int_{-\infty }^\infty d\nu
_p\int_{-\infty }^\infty d\nu _m\cdot |g(\nu _p)|^2e^{-(\nu _p^2+\nu
_m^2)/(2\sigma ^2)}  \nonumber \\
&=&\sqrt{2\pi }\sigma \int_{-\infty }^\infty d\nu _p\cdot |g(\nu
_p)|^2e^{-\nu _p^2/(2\sigma ^2)}.  \nonumber
\end{eqnarray}
Using Eq. (\ref{29t}), we calculate the coincidence probability 
\begin{eqnarray}
P_c &=&\frac 12\{1-\int_{-\infty }^\infty d\omega _1\int_{-\infty }^\infty
d\omega _2[C_{\alpha \beta }(\omega _1,\omega _2)C_{\beta \alpha
}^{*}(\omega _2,\omega _1)+\text{c.c.}]\}  \eqnum{D2}  \label{D2} \\
&=&\frac 12\{1-\int_{-\infty }^\infty d\nu _1\int_{-\infty }^\infty d\nu
_2\cdot |g(\nu _1+\nu _2)|^2e^{-(\nu _1^2+\nu _2^2)/\sigma ^2}2\cos [(\nu
_2-\nu _1+\Omega _\beta -\Omega _\alpha )\frac{\Delta z}c-\theta ]\} 
\nonumber \\
&=&\frac 12\{1-\int_{-\infty }^\infty d\nu _p\int_{-\infty }^\infty d\nu
_m\cdot |g(\nu _p)|^2e^{-(\nu _p^2+\nu _m^2)/(2\sigma ^2)}\cos [(\nu
_m+\Omega _\beta -\Omega _\alpha )\frac{\Delta z}c-\theta ]\}  \nonumber \\
&=&\frac 12\{1-\int_{-\infty }^\infty d\nu _p|g(\nu _p)|^2e^{-\nu
_p^2/(2\sigma ^2)}\int_{-\infty }^\infty d\nu _m\cdot e^{-\nu _m^2/(2\sigma
^2)}\cos [(\nu _m+\Omega _\beta -\Omega _\alpha )\frac{\Delta z}c-\theta ]\}
\nonumber \\
&=&\frac 12\{1-\int_{-\infty }^\infty |g(\nu _p)|^2e^{-\nu _p^2/(2\sigma
^2)}d\nu _p\cdot \sqrt{2\pi }\sigma \cos [(\Omega _\beta -\Omega _\alpha )%
\frac{\Delta z}c-\theta ]e^{-\frac 12\Delta z(\sigma /c)^2}\}.  \nonumber
\end{eqnarray}
By taking into account the normalization (\ref{D1}), one obtains Eq. (\ref
{57}).

Second, we consider the case of $\sigma _\alpha \neq \sigma _\beta $. For
the spectrum described by Eqs. (\ref{13}) and (\ref{58}), the normalization
is given by 
\begin{eqnarray}
1 &=&2\int_{-\infty }^\infty d\omega _1\int_{-\infty }^\infty d\omega
_2|C_{\alpha \beta }(\omega _1,\omega _2)|^2=2A^2\int_{-\infty }^\infty d\nu
_1\int_{-\infty }^\infty d\nu _2\cdot e^{-(\nu _1+\nu _2)^2/\sigma _p^2-\nu
_1^2/\sigma _\alpha ^2-\nu _2^2/\sigma _\beta ^2}  \eqnum{D3}  \label{D3} \\
&=&A^2\int_{-\infty }^\infty d\nu _p\int_{-\infty }^\infty d\nu _m\cdot
e^{-\nu _p^2/\sigma _p^2-(\nu _p+\nu _m)^2/(4\sigma _\alpha ^2)-(\nu _p-\nu
_m)^2/(4\sigma _\beta ^2)}  \nonumber \\
&=&2A^2\pi \frac{\sigma _p\sigma _\alpha \sigma _\beta }{\sqrt{\sigma
_p^2+\sigma _\alpha ^2+\sigma _\beta ^2}}.  \nonumber
\end{eqnarray}
The coincidence probability is calculated as 
\begin{eqnarray}
P_c &=&\frac 12\{1-A^2\int_{-\infty }^\infty d\nu _1\int_{-\infty }^\infty
d\nu _2\cdot e^{-(\nu _1+\nu _2)^2/\sigma _p^2-\nu _1^2/\sigma _\alpha
^2-\nu _2^2/\sigma _\beta ^2}2\cos [(\nu _2-\nu _1)\frac{\Delta z}c+(\Omega
_\beta -\Omega _\alpha )\frac{\Delta z}c-\theta ]\}  \eqnum{D4}  \label{D4}
\\
&=&\frac 12\{1-A^2\int_{-\infty }^\infty d\nu _p\int_{-\infty }^\infty d\nu
_m\cdot e^{-\nu _p^2/\sigma _p^2-(\nu _p+\nu _m)^2/(4\sigma _\alpha ^2)-(\nu
_p-\nu _m)^2/(4\sigma _\beta ^2)}\cos [\nu _m\frac{\Delta z}c+(\Omega _\beta
-\Omega _\alpha )\frac{\Delta z}c-\theta ]\}  \nonumber \\
&=&\frac 12\{1-2A^2\pi \frac{\sigma _p\sigma _\alpha \sigma _\beta }{\sqrt{%
\sigma _p^2+\sigma _\alpha ^2+\sigma _\beta ^2}}\cos [(\Omega _\beta -\Omega
_\alpha )\frac{\Delta z}c-\theta ]\cdot e^{-\frac 12\Delta z^2(\sigma
/c)^2}\},  \nonumber
\end{eqnarray}
where $\sigma $ is defined by Eq. (\ref{61}). By taking into account the
normalization (\ref{D3}), one obtains Eq. (\ref{57}) again.

\bigskip\ captions of figures

Fig. 1 Spectral plane for the input state. (a) There is not a pair of
two-photon states to be interfered because $|C(\omega _1,\omega _2)C(\omega
_2,\omega _1)|=0$; (b) There are pairs of symmetric two-photon states and
the degenerate two-photon states which may interfere.

Fig. 2 Contours of two-photon spectra for two independent single-photon
wavepackets, each of which is described by a Gaussian-type. (a) two
wavepackets have the same central frequency, and (b) two wavepackets have
different central frequencies.

Fig. 3 Coincidence probability versus phase $2\theta $ at the balanced
position $z_1=z_2$ for the perfect entanglement $g(x)=\delta (x)$ (solid
line), the arbitrary entanglement $g(x)$ defined by Eq. (\ref{13}) with $%
\sigma _p=\sigma $ (dashed line), and two independent single-photon
wavepackets (dotted line). The normalized optical path difference: (a) $%
\Delta L(\sigma /c)=1$, (b) $\Delta L(\sigma /c)=3$.

Fig. 4 Coincidence probability versus the normalized position of
beamsplitter, $\Delta z(\sigma /c)$, for (a) $\theta =(n+1/2)\pi $, (b) $%
\theta =n\pi $, and (c) $\theta =(n+1/4)\pi $. Other illustrations are the
same as Fig. 3.

Fig. 5 Same as in Fig. 4 but $\Delta L(\sigma /c)=3$.

Fig. 6 Contours of un-entangled two-photon spectra for (a) $\Delta L(\sigma
/c)=1$ and $\theta =n\pi ;$ (b) $\Delta L(\sigma /c)=1$ and $\theta
=(n+1/2)\pi ;$ (c) $\Delta L(\sigma /c)=3$ and $\theta =n\pi ;$ (d) $\Delta
L(\sigma /c)=3$ and $\theta =(n+1/2)\pi $.

Fig. 7 Contours of entangled two-photon spectra with $\sigma _p=(1/3)\sigma $%
; other parameters are the same as in Fig. 6.

Fig. 8 Total coincidence probability versus phase $\theta $ at the balanced
position $z_1=z_2$ for two independent single-photon wavepackets being in
two orthogonal polarizations (curve 1) and the polarization entangled
two-photon wavepacket (curve 2). For the former, $n_\alpha =n_\beta =1/2$ is
set.

Fig. 9 For the polarization entangled two-photon wavepacket with $\Omega
_\alpha =\Omega _\beta $, the coincidence probabilities versus the
normalized position of beamsplitter, $\Delta z(\sigma /c),$ for $\theta =0$, 
$\pi /4$, $\pi /2$, $3\pi /4$, and $\pi $.

Fig. 10 Same as in Fig. 9 but $(\Omega _\beta -\Omega _\alpha )/(\pi \sigma
)=2$ and (a) $\theta =0$, (b) $\theta =\pi /2$, (c) $\theta =\pi $.

Fig. 11 For two independent single-photon wavepackets being in two
orthogonal polarizations with $n_\alpha =n_\beta =1/2,$ $\Omega _\alpha
=\Omega _\beta $ and $\sigma _\alpha =\sigma _\beta \equiv \sigma $,
coincidence probabilities versus the normalized position of beamsplitter, $%
\Delta z(\sigma /c)$, for (a) $\theta =0$, (b) $\theta =\pi /2$, and (c) $%
\theta =\pi $. Dotted, dashed and solid curves are for the coincidence
probabilities of the same polarization photons $P_c^{\alpha \alpha
}(P_c^{\beta \beta }),$ the cross-coincidence probability $P_c^{\alpha \beta
}+P_c^{\beta \alpha }$ and the total coincidence probability $P_c$,
respectively.

Fig. 12 Same as in Fig. 11 but $(\Omega _\beta -\Omega _\alpha )/(\pi \sigma
)=2$ and (a) $\theta =0$, (b) $\theta =\pi /2$, (c) $\theta =\pi $.

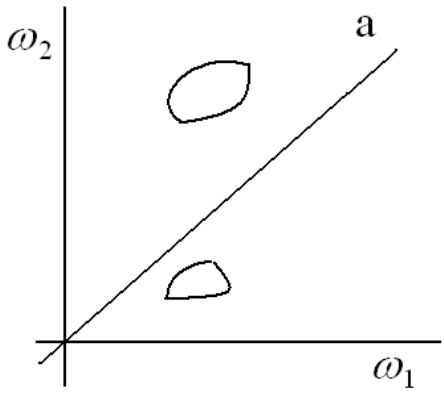
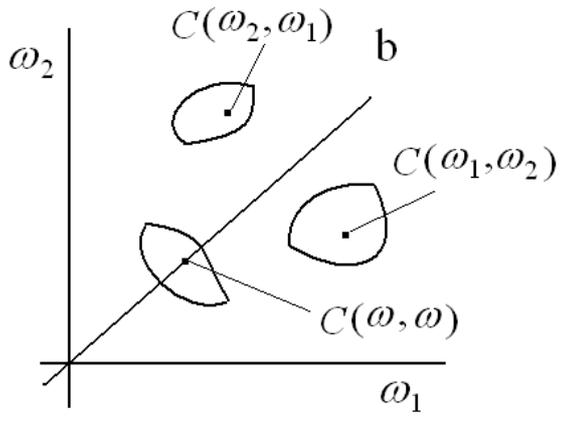
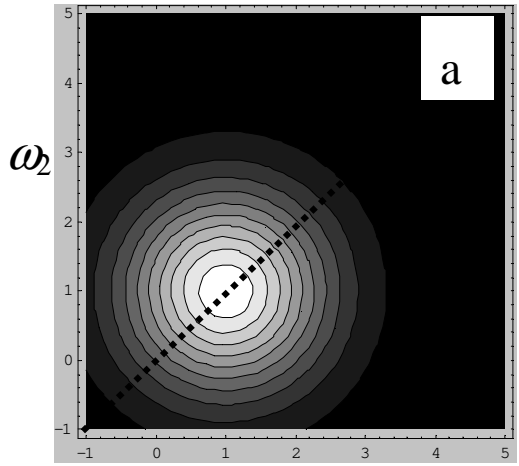
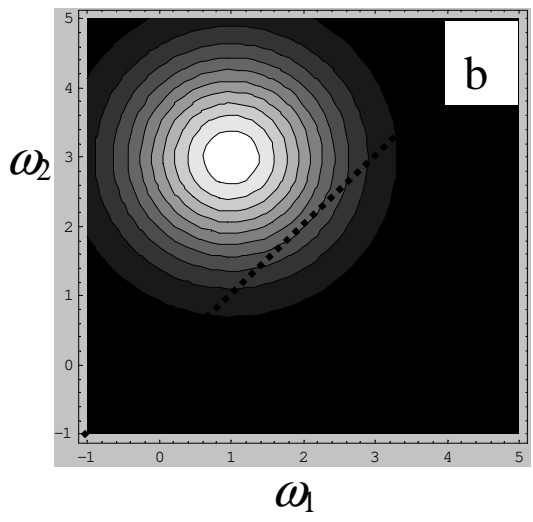
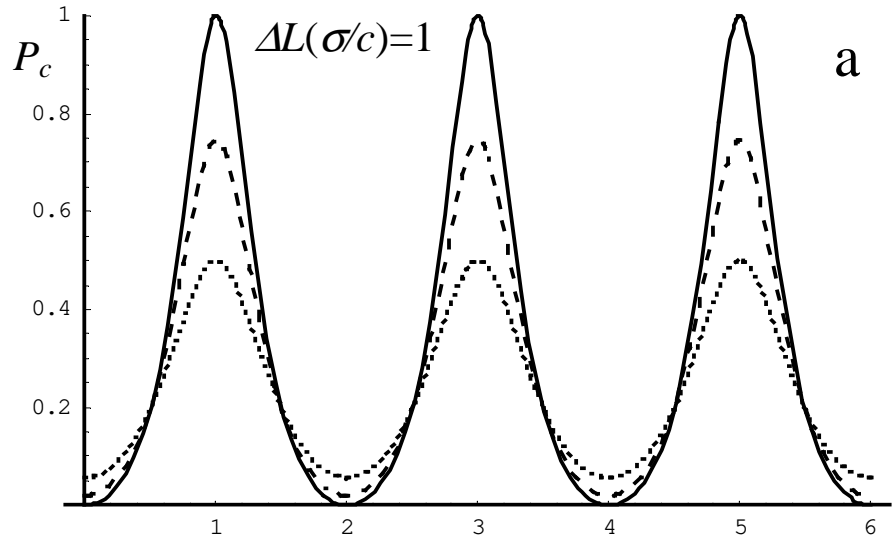
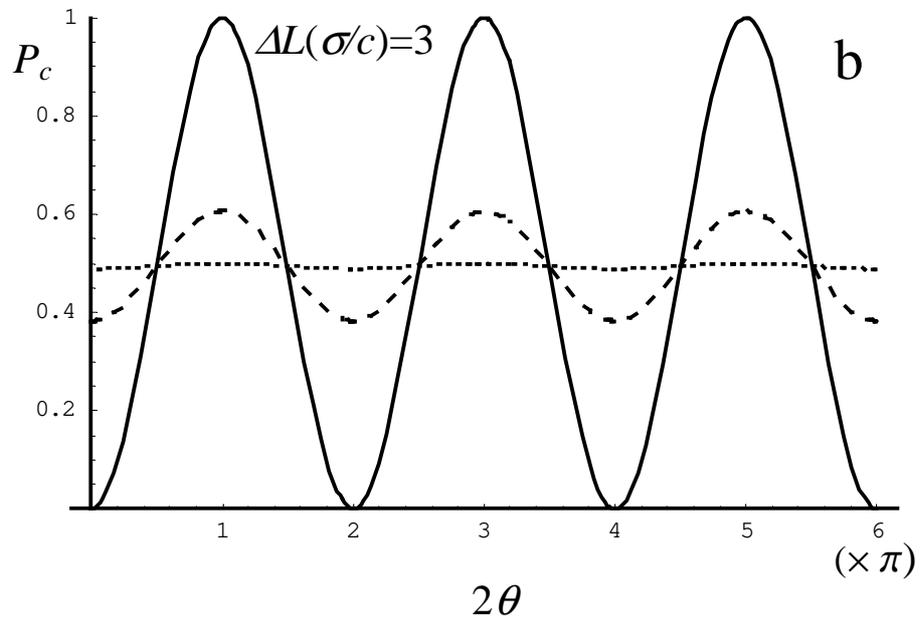
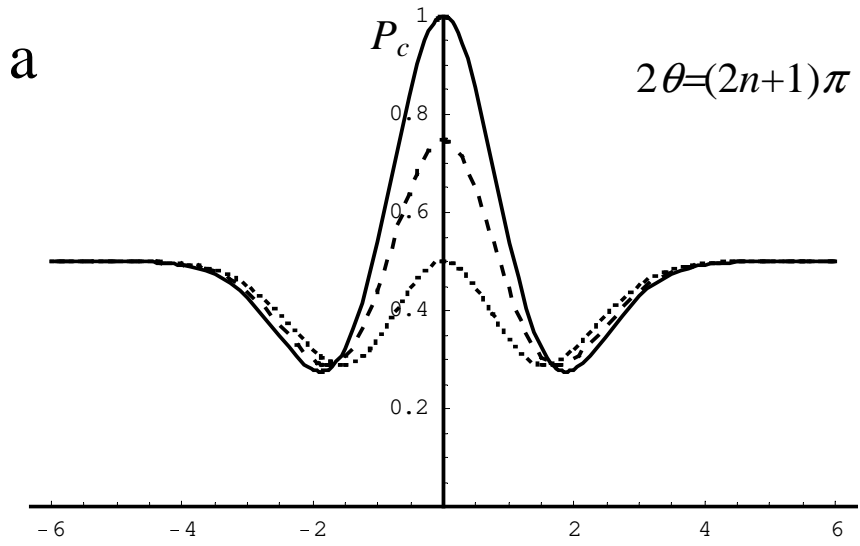
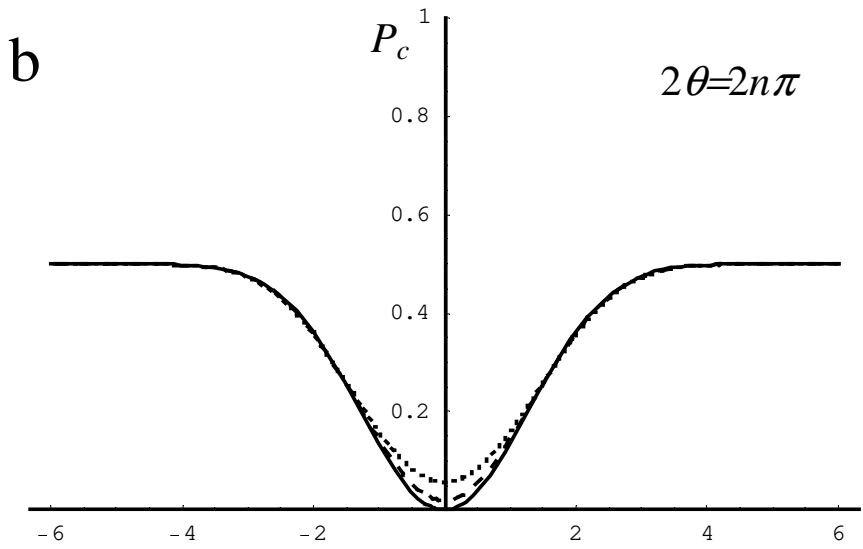
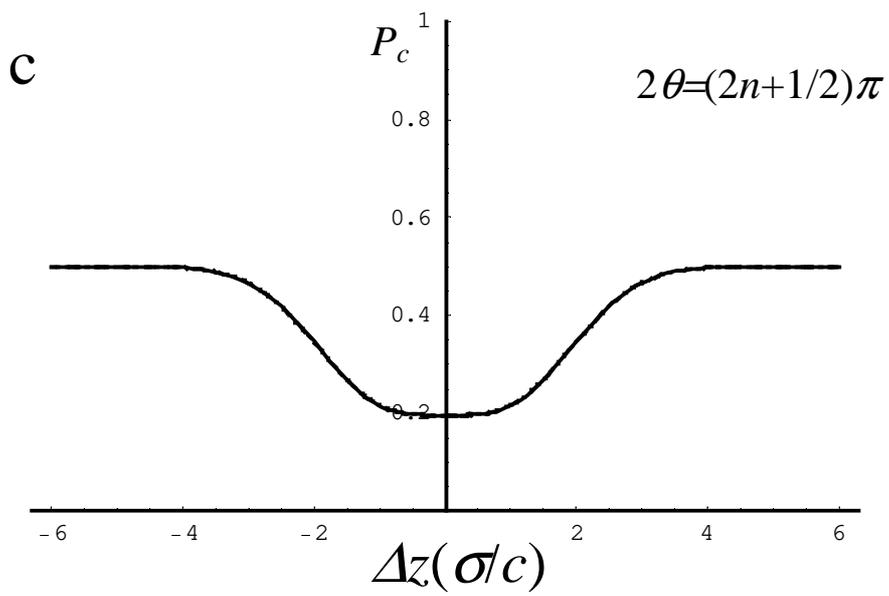
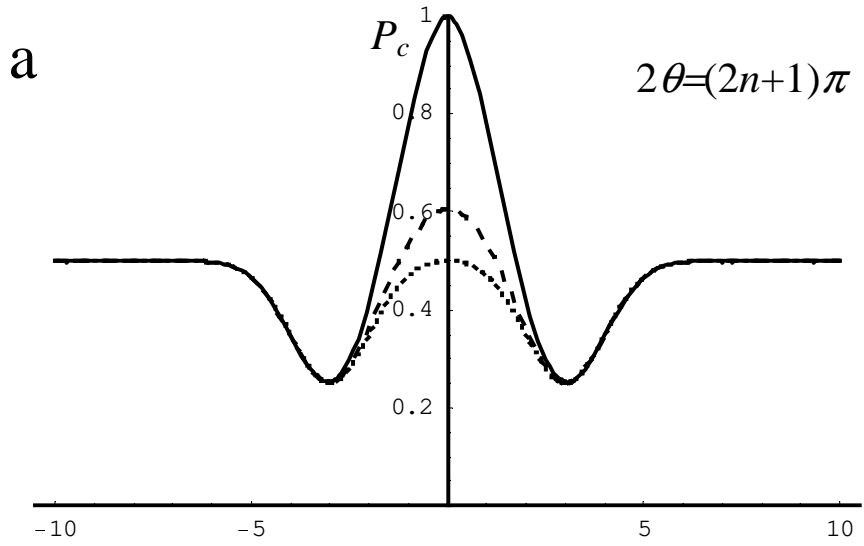
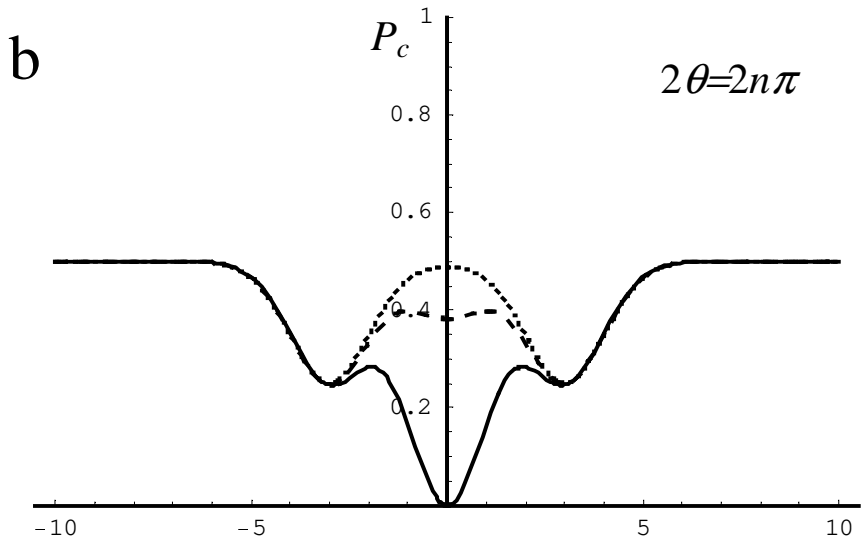
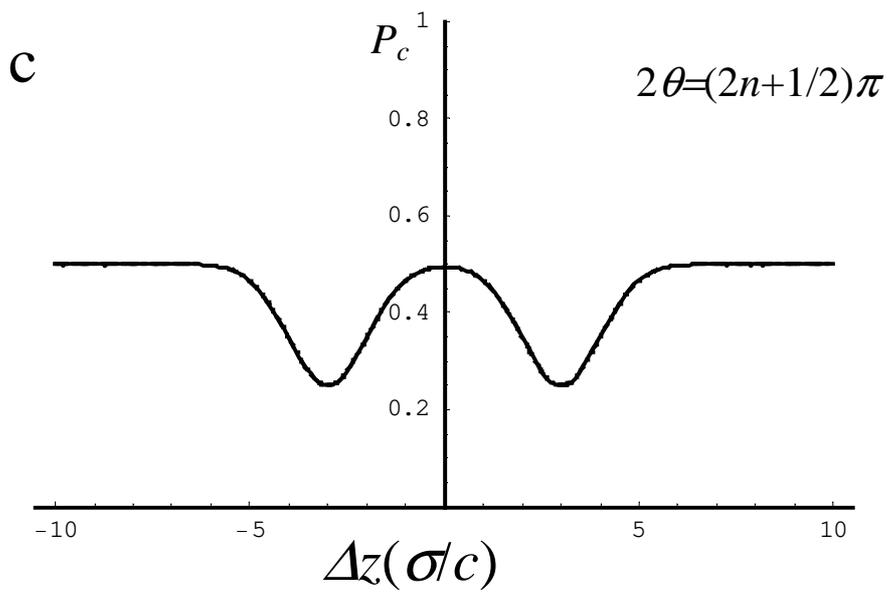
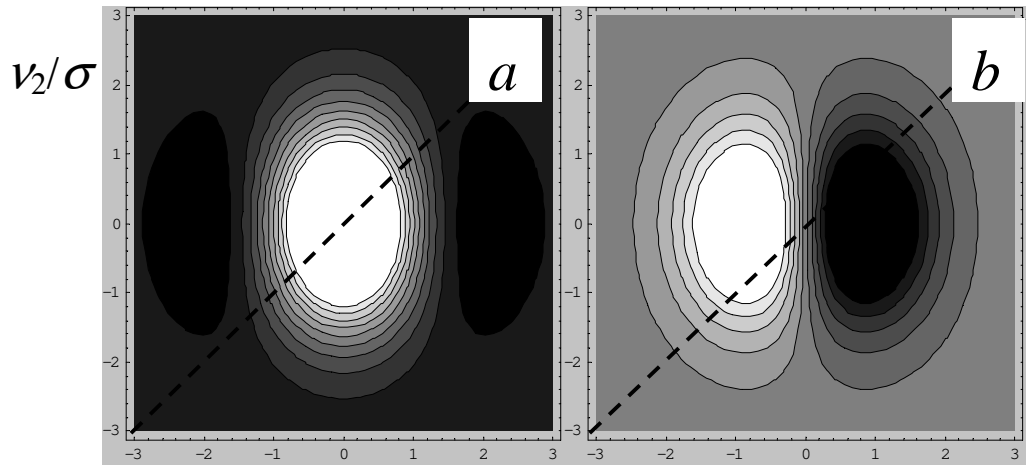
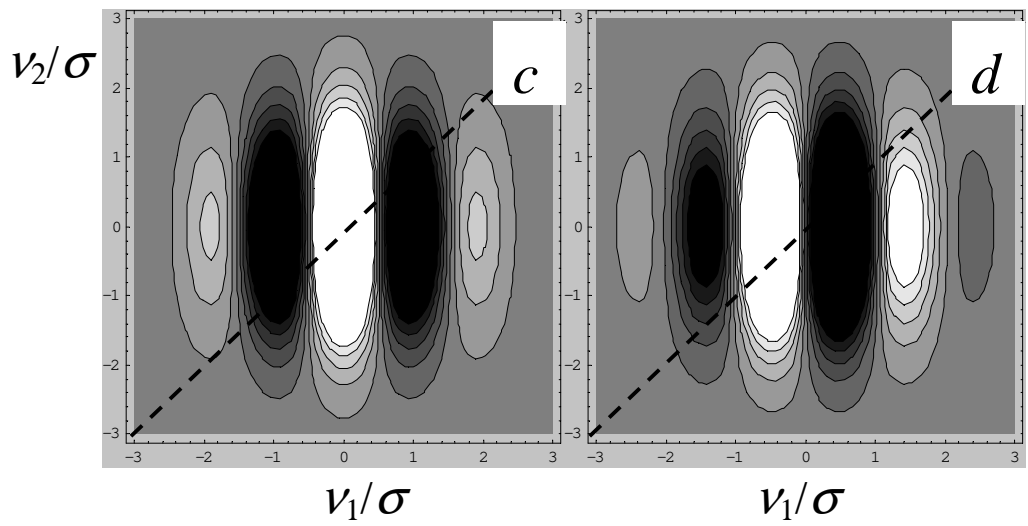
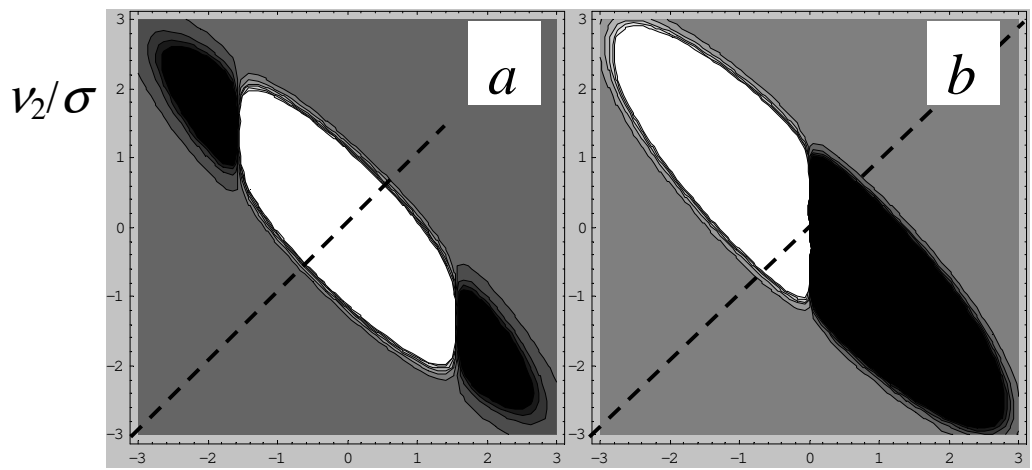
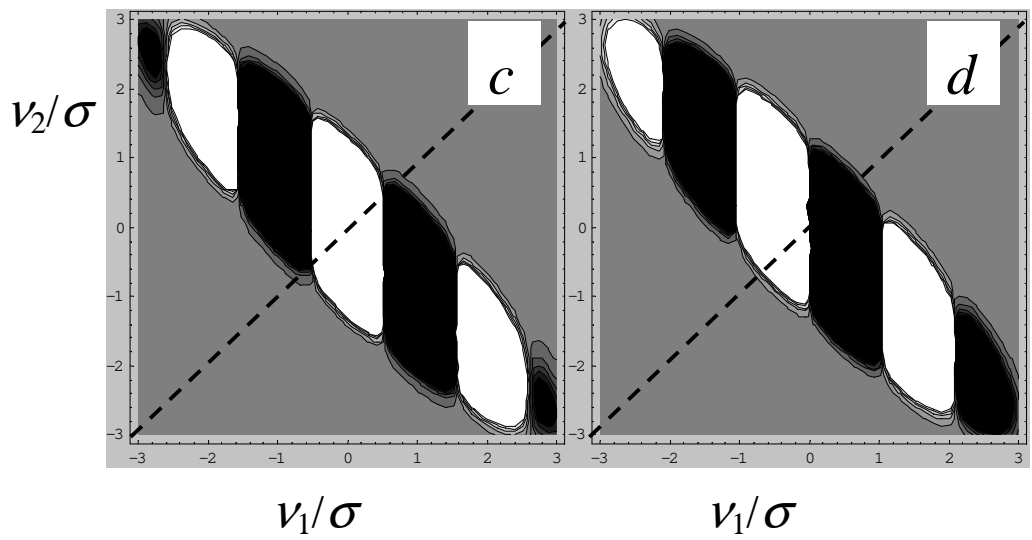
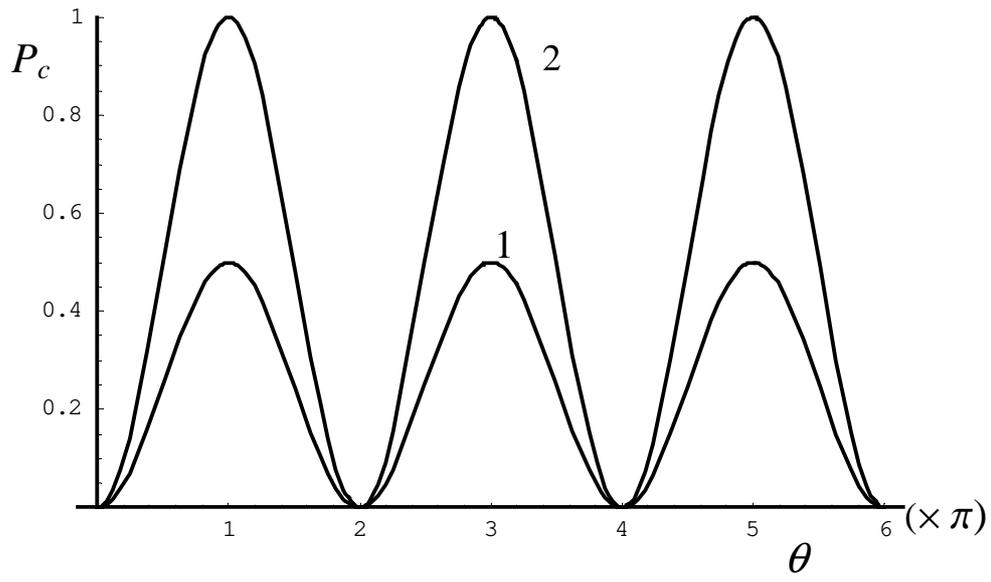
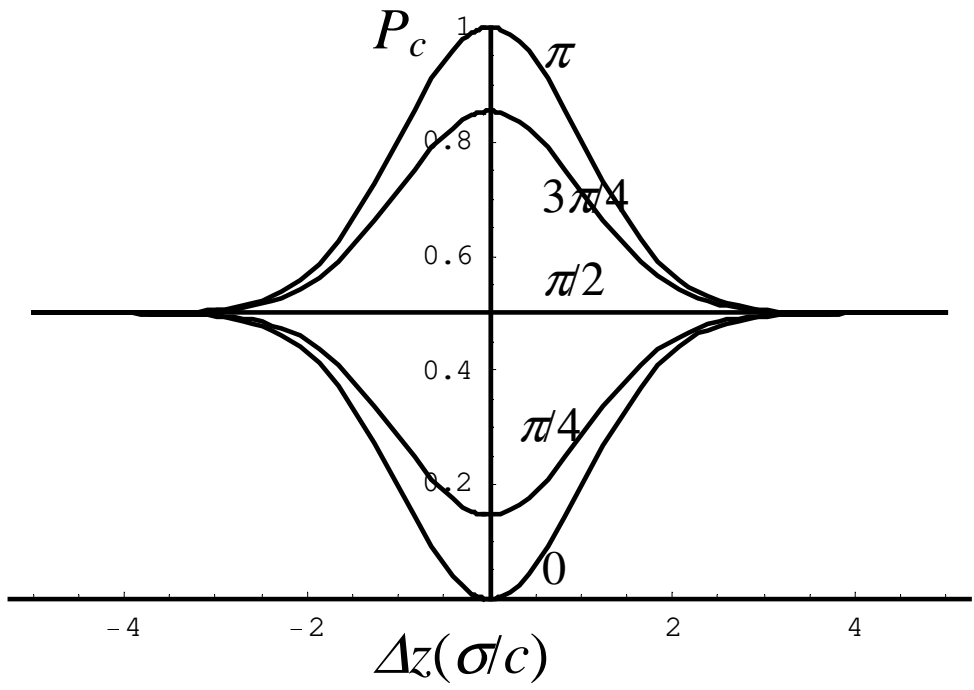
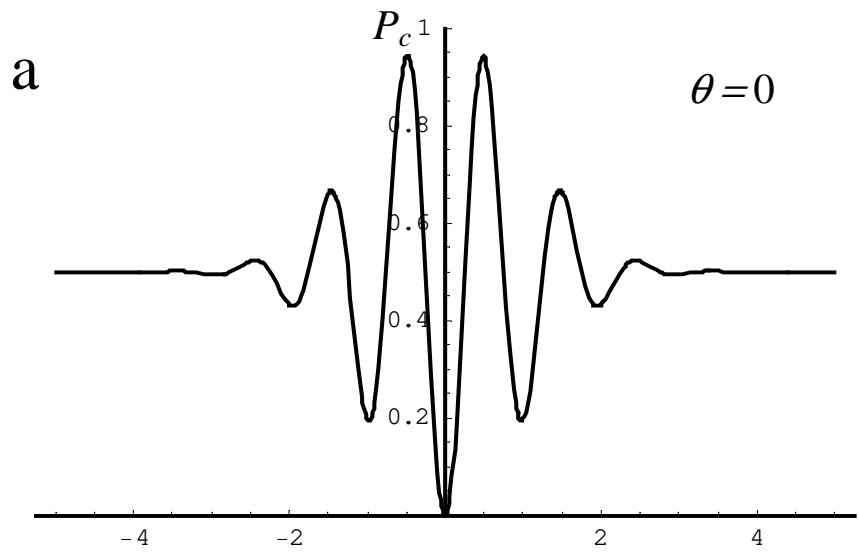
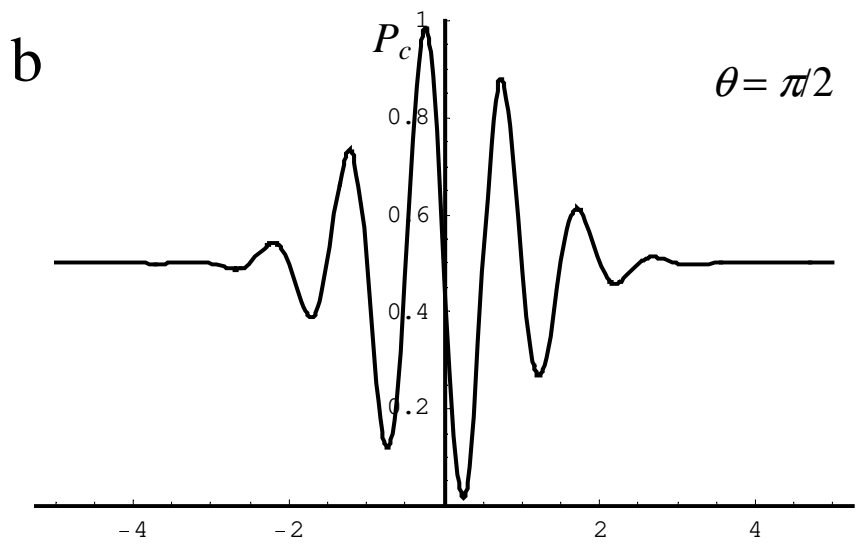
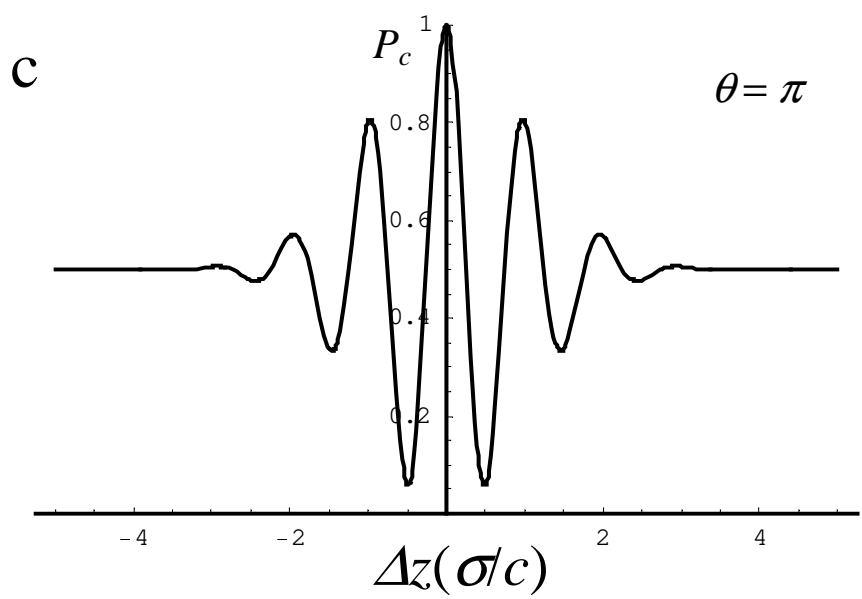
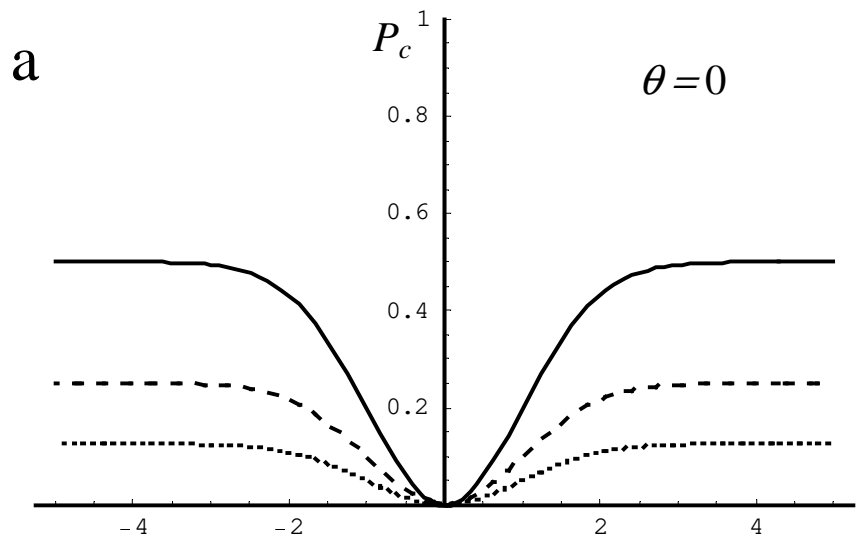
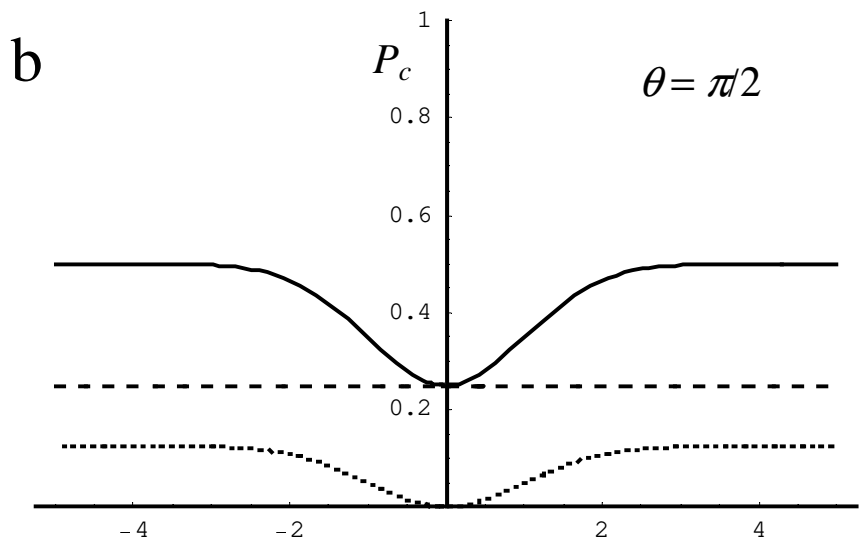
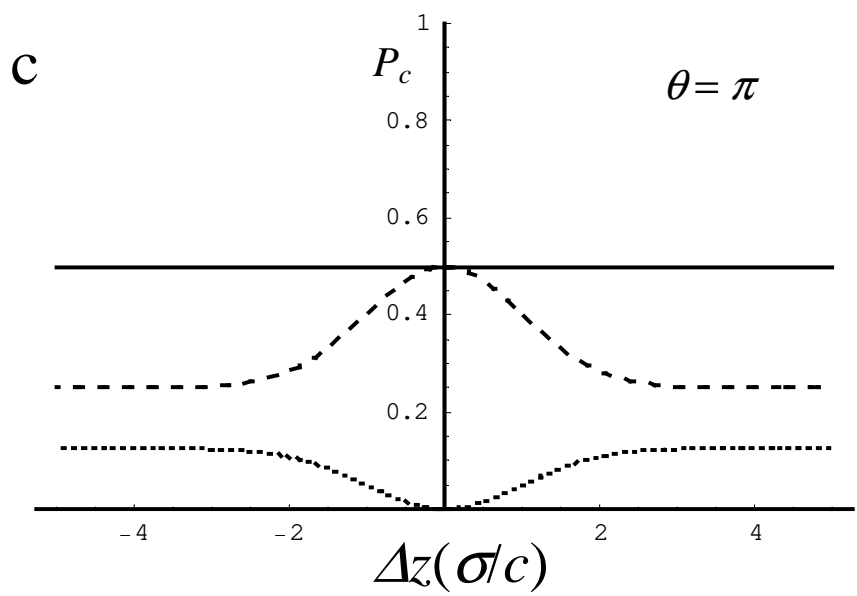
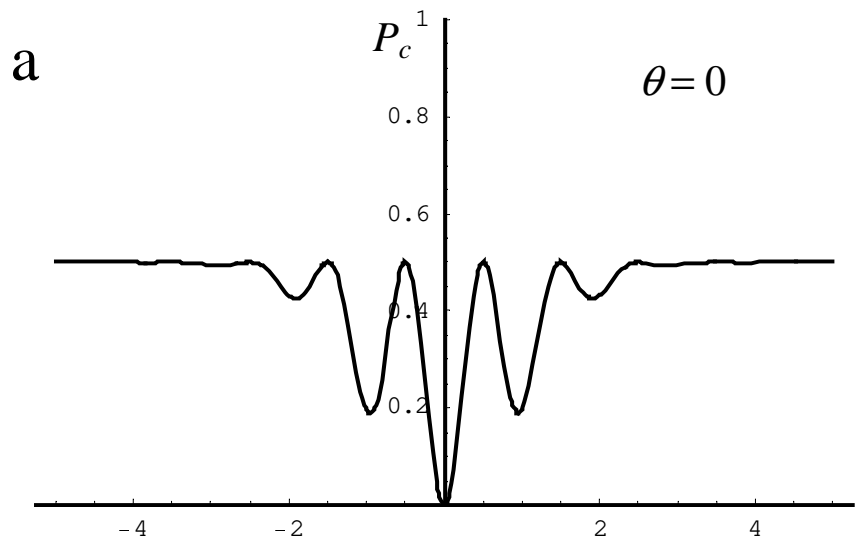
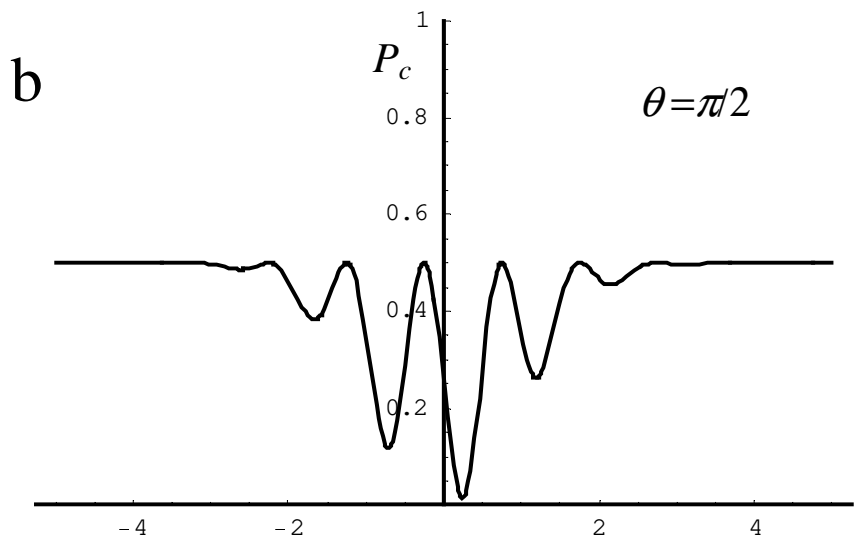
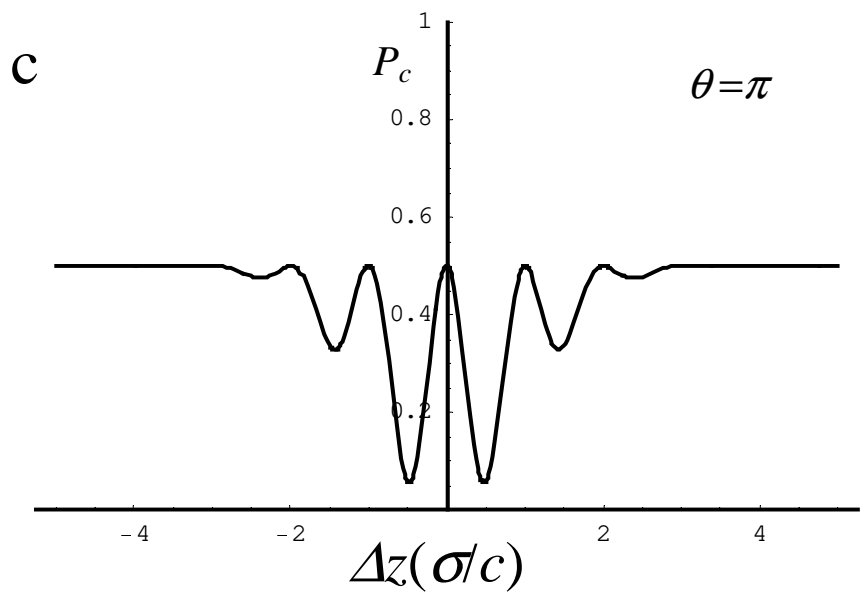
\end{document}